\documentclass[a4paper,fleqn]{cas-sc}

\usepackage[authoryear,longnamesfirst]{natbib}
\usepackage{graphicx}
\usepackage{subcaption}
\usepackage{float}
\usepackage{amsthm}
\usepackage{algorithm}
\usepackage{algpseudocode}
\usepackage{tikz}
\usepackage{booktabs}
\usepackage{subcaption} 

\def\tsc#1{\csdef{#1}{\textsc{\lowercase{#1}}\xspace}}
\tsc{WGM}
\tsc{QE}

\theoremstyle{definition}
\newtheorem{definition}{Definition} %

\begin{document}
\let\WriteBookmarks\relax
\def\floatpagepagefraction{1}
\def\textpagefraction{.001}

\shorttitle{}    

\shortauthors{}  

\title [mode = title]{Spatiotemporal Dynamics of Conway's Game of Life in the Entropy–Complexity Plane}

\author[1]{Lucas Malacarne Astore}[orcid=0009-0003-5541-8340]
\cormark[1]
\ead{lucasastore@ufmg.br}

\author[2]{Alejandro C. Frery}[orcid=0000-0002-8002-5341]

\author[3]{Petrônio Cândido de Lima e Silva}[orcid=0000-0002-1202-2552]

\author[1]{Frederico Gadelha Guimarães}[orcid=0000-0001-9238-8839]

\affiliation[1]{organization={Federal University of Minas Gerais},
            city={Belo Horizonte},
            state={Minas Gerais},
            country={Brazil}}

\affiliation[2]{organization={Victoria University of Wellington},
            city={Wellington},
            country={New Zealand}}

\affiliation[3]{organization={Federal Institute of Northern Minas Gerais},
            state={Minas Gerais},
            country={Brazil}}

\cortext[1]{Corresponding author}

\begin{abstract}
We investigate the dynamics of 
Conway's Game of Life (GoL) through the lens of Ordinal Pattern analysis. 
Each GoL generation is transformed into a one-dimensional time series, from which ordinal patterns are extracted to characterize its underlying dynamics. 
The resulting ordinal distributions are then represented in the Entropy--Statistical Complexity ($H \times C$) plane, enabling a quantitative analysis of the system's evolution. 
We evaluate different encoding strategies, including patch sizes and linearization methods, and assess their impact on the resulting $H \times C$ 
trajectories. 
Finally, we characterize the emergent dynamics in terms of 
convergence behavior, 
trajectory patterns, 
and statistical complexity, 
showing how the proposed entropy--complexity framework provides insights into the dynamics of cellular automata.
\end{abstract}

\begin{highlights}
\item Ordinal patterns reveal the dynamics of the Game of Life.
\item Different initial densities lead to distinct but convergent steady states.
\item The framework captures transient and long-term spatiotemporal organization.
\item Peano-Hilbert curve encoding maps Game of Life dynamics onto the $H\times C$ plane for spatiotemporal complexity analysis
\item Exponential relaxation model quantifies convergence rate and half-life, validated across lattice size and bit-depth
\end{highlights}

\begin{keywords}
Cellular Automata \sep 
entropy--complexity Plane \sep 
Ordinal Patterns \sep 
Bandt-Pompe method \sep 
Peano-Hilbert curve
\end{keywords}

\maketitle

\section{Introduction}\label{intro}

Complexity in dynamical systems can be defined as the interaction of unit components, which can give rise to a wide variety of collective behaviors. 
Rather than being determined solely by the complexity of individual components, such behavior can arise from their local interactions and the resulting organization of the system over time. 
Cellular automata provide classical examples of a complex dynamical systems: despite being governed by 
simple, 
local, and 
deterministic transition rules, they can generate intricate spatiotemporal patterns whose long-term behavior is often computationally irreducible or unpredictable.

According to \citet{Wolfram1983Statistical}, 
cellular automata (CAs) were formally introduced through the work of John von Neumann and Stanis\l{}aw 
Ulam in the 1940s, initially in the context of studying self-reproduction and the organization of computational systems. 
The term \emph{cellular} reflects the decomposition of the system into discrete, 
locally interacting units, 
providing a simple computational framework in which global structures may emerge from local interactions \citep{ConwaysGameo2024l}. 
Since then, cellular automata have been employed as models for a broad range of dynamical and physical phenomena, providing a convenient setting for investigating emergence of 
macroscopic complex behavior, 
self-organization, and 
information propagation.

Among CA setups, Conway's Game of Life (GoL) has become one of the most extensively studied rules \citep{Conway19704,Speck1978,ConwaysGameo2024l,johnston2021conway}. 
Despite its simple local update rules, GoL generates a rich variety of spatiotemporal structures, including stable configurations, oscillators, moving structures, and interacting patterns. 
It is commonly associated with Wolfram's Class~IV CA classification, characterised by localized structures, long-range interactions, and the propagation and processing of information \citep{Wolfram2002}. 
These properties make GoL an especially informative test case for the study of complexity: 
unlike highly ordered systems that rapidly converge to simple configurations, or highly disordered systems that exhibit predominantly stochastic behavior, GoL can sustain non-trivial organization 
over long periods of time.

An important challenge in the analysis of such systems is therefore to quantify not only their degree of disorder, but also the structural organization underlying their dynamics. 
Entropy provides a natural measure for quantifying uncertainty or disorder. 
In physical systems, for example, entropy is associated with the distribution of accessible states, while dynamical processes can be understood in terms of  
how information and system states change across space and time.
Higher entropy corresponds to a greater diversity of possible states, whereas lower entropy indicates a more constrained organization. 
However, entropy alone does not distinguish between different types of organization that may exhibit similar levels of uncertainty.

This limitation is particularly relevant for cellular automata. 
Interestingly, whereas physical systems are often associated with the tendency toward increasingly disordered configurations, 
cellular automata can generate highly organized structures from initially disordered states through deterministic local interactions \citep{d2015classification}. 
Consequently, characterizing CA dynamics solely through entropy may overlook important aspects of their structural organization. 
Information-theoretic approaches that jointly consider uncertainty and complexity can provide a richer description of these dynamics. 
In particular, 
the entropy--complexity ($H\times C$) approach
provides a framework in which entropy characterizes the degree of uncertainty while
statistical complexity captures structural organization beyond randomness.

Several studies have investigated cellular automata from an information-theoretic perspective. 
Early approaches considered entropy-related properties of cellular-automaton dynamics \citep{Speck1978}, 
while more recent studies have explored entropy and information-based measures to characterize emergent behavior in CA systems \citep{pan2010emergencesymmetrynewtype, ContrerasReyes2021Lerch}. 
\citet{ContrerasReyes2021Lerch}, for example, connected GoL dynamics with information-theoretic measures by characterizing the distribution of living-cell populations through the maximum nonsymmetric entropy principle. 
These studies show that information-based measures can provide meaningful descriptions of GoL dynamics. 
Nevertheless, much of the existing literature characterizes the system through global quantities such as the population or density of living cells, rather than directly quantifying the informational organization of the spatial configuration itself.

This distinction is important because the same proportion of living cells can correspond to substantially different spatial organizations. 
Two GoL configurations may therefore exhibit similar population densities while presenting very different arrangements, structures, and interaction patterns. 
A measure capable of capturing the organization encoded in the spatial configuration can consequently provide complementary information to population-based entropy measures.

In this work, this issue was addressed by representing each two-dimensional 
GoL generation 
as a one-dimensional sequence while preserving, as much as possible, the spatial relationships among neighboring cells. 
To this end, the Peano-Hilbert (PH) space-filling curve was employed, which provides a locality-preserving traversal of a two-dimensional lattice: cells that are spatially close tend to remain close in the resulting one-dimensional ordering. 
This transformation allows the spatial configuration of each generation to be represented as a one-dimensional 
sequence suitable for time-series-based information-theoretic analysis.

\begin{figure}[]
\centering
\includegraphics[width=0.8\linewidth]{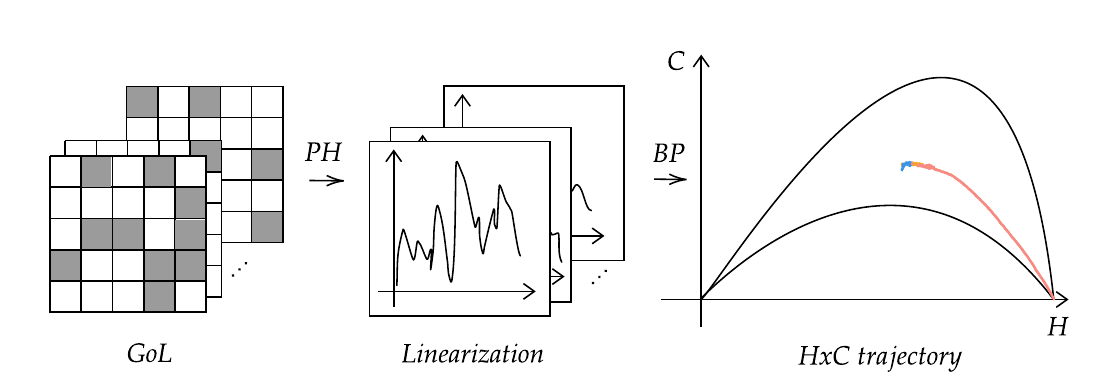}
\caption{Overview of the proposed approach. The spatiotemporal dynamics of the Game of Life cellular automaton (CA) are first linearized into a one-dimensional sequence using a Peano-Hilbert (PH) space-filling curve. 
The resulting sequence is then mapped onto the entropy--complexity plane using Bandt-Pompe~(BP) ordinal patterns. 
Each CA generation is consequently characterized in terms of its entropy and statistical complexity, and the sequence of generations traces a trajectory in the entropy--complexity plane.}
\label{fig:intro}
\end{figure}

The resulting sequence is subsequently transformed into ordinal patterns \citep{bandt2002permutation}, 
and from their histogram we compute the permutation entropy and statistical complexity. 
Each generation of the CA can therefore be represented as a point in the $H\times C$ plane, 
allowing the evolution of the informational organization of the spatial configuration to be followed throughout the CA dynamics. 
Figure~\ref{fig:intro} illustrates our methodological approach.

The central research question addressed in this work is: 
\emph{How does the informational organization of the spatial configuration of Conway's Game of Life evolve over time?} 
More specifically, we investigate the long-term global dynamics of GoL through the evolution of its entropy and statistical complexity, seeking to characterize how the spatial organization of the CA changes throughout its trajectory.

The rationale for focusing on GoL is also methodological. 
As emphasized by \citet{johnston2021conway}, GoL is arguably the most extensively studied cellular-automaton rule. 
Rather than introducing another rule and obtaining only a limited characterization of its behavior, we follow the complementary strategy of asking how far an information-theoretic analysis can be pushed when applied to a well-established and extensively studied system. 
This allows the proposed representation to be evaluated in a setting whose dynamical properties have been extensively documented.

To the best of our knowledge, this is the first study to characterize the spatiotemporal evolution of Conway's Game of Life directly in the entropy--complexity plane by encoding its two-dimensional spatial configurations through a Peano-Hilbert traversal and analyzing the resulting sequences using ordinal patterns. 
The proposed framework provides a representation of the spatiotemporal organization of GoL that complements existing population-based entropy analyses and enables the long-term evolution of its informational structure to be investigated through trajectories in the $H\times C$ plane.

\section{Cellular Automata}

The concept of cellular automaton (CA) was introduced by John von Neumann in 1947 as a discrete computational framework for investigating self-reproduction and self-repair in biological systems and their potential application to electronic circuits \citep{neumann,Chopardbook}. In collaboration with Stanis\l{}aw Ulam, von Neumann reformulated the original continuous model as a lattice of interconnected finite-state cells, yielding a mathematically tractable framework for studying complex dynamical systems \citep{KARI20053}. 

A CA consists of a lattice of cells, each assuming one of a finite number of possible states at time $t$. The state of every cell is updated synchronously according to a local transition rule that depends on the states of its neighboring cells. Starting from an initial configuration, the repeated application of these local rules generates the spatiotemporal dynamics of the system in discrete space and time, often giving rise to complex emergent behavior from simple microscopic interactions. 
Expanding from \citet{KARI20053} and \citet{Rollier_2025}, we define a CA as a quadruple.
\begin{definition}[Cellular automaton] \label{def:CA}
A cellular automaton (CA) is a quadruple $ \left\langle \mathbb{Z}^d, S, \Delta, \phi \right\rangle$, where:
\begin{enumerate}
    \item The CA is defined on the countably infinite $d$-dimensional \emph{lattice} of cells $\mathbb{Z}^d$; common cases are one-, two-, and three-dimensional grids ($d=1,2,3$). %

    \item A \emph{configuration} at time $t$ is a function $c_{t}\colon \mathbb Z^d \rightarrow S$, $t=\{0,1,\dots\}$, that assigns a state from a finite \emph{state set} $S$  to every cell in the lattice. 
    The set of all possible configurations $\mathcal{C} = S^{\mathbb{Z}^d}$ is the \emph{configuration space}.
    For every $A\subset\mathbb Z^d$, we denote the observed states in $A$ at time $t$ as $c_t(A)$.

    \item  For every $x\in\mathbb Z^d$, $\partial_x \subset \mathbb{Z}^{d}\setminus\{x\}$ is a finite neighborhood of $x$, usually defined by means of a distance between points in $\mathbb{Z}^d$.
   The closure of site $x$ is $\overline{\partial}_x=\partial_x\cup\{x\}.$

    \item The set of all closures in a lattice are denoted by $\Delta = \{ \overline{\partial}_{x} : x \in \mathbb{Z}^d\}$

     \item $\phi: \Delta \rightarrow S$ is the \emph{local update rule},
         determining the next state of a cell from the current states of its neighborhood:
        \begin{equation}
            c_{t+1}(x) = \phi\big(c_t(\overline{\partial}_x)\big).
            \label{eq:TransitionFunction}
        \end{equation}
        
\end{enumerate}

The local rule $\phi$ induces a \emph{global transition map} $F\colon \mathbb{Z}^d \rightarrow \mathbb{Z}^d$ defined cell-wise, so that the dynamics of the CA, given an initial  configuration $c_0$:
\begin{equation}
    c_{t+1} = F\big(c_t\big), \qquad t = 0, 1, 2, \dots .
    \label{eq:GlobalDynamics}
\end{equation}

\end{definition}

Given this general framework, a wide range of cellular automaton can be constructed, ranging from one-dimensional systems, such as the well-known elementary cellular automata (ECA) with specific emergent patterns \cite{KARI20053}, to higher-dimensional systems with multiple states per cell. 

A defining characteristic of cellular automata is their ability to generate complex spatiotemporal structures from simple local interaction rules. 
Despite being governed by deterministic and precisely defined transition functions, cellular automata can exhibit a rich variety of dynamical behaviors, including stable configurations, periodic oscillations, and chaotic patterns.

During the 1980s, Stephen Wolfram extensively investigated the dynamical behavior of cellular automata, introducing a qualitative classification scheme based on the behavior complexity of their spatiotemporal dynamics. 
This taxonomy, later summarized in his book ``A New Kind of Science'' \citep{Wolfram2002}, categorizes cellular automata into four main classes:
\begin{itemize}
\item \textbf{Class I (Homogeneous behavior).} Cellular automata in this class evolve toward a uniform configuration in which all cells eventually reach a stable state.

\item \textbf{Class II (Periodic behavior).} Systems in this class evolve into simple periodic structures, including fixed points and oscillatory patterns that repeat indefinitely.

\item \textbf{Class III (Chaotic behavior).} Cellular automata in this class exhibit highly irregular and aperiodic spatiotemporal patterns, characterized by sensitivity to initial conditions and random-like behavior.

\item \textbf{Class IV (Complex behavior).} Cellular automata in this class generate localized structures and long-range interactions, producing complex patterns that combine ordered and chaotic features. These systems can support persistent structures, information propagation, and computationally universal behavior. The classical Conway's Game of Life belongs to this class and is discussed in the following section.

\end{itemize}

\subsection{Conway's Game of Life}

Among the many cellular automata proposed in the literature, Conway's Game of Life (GoL), introduced by \citet{Conway19704}, is arguably the most widely studied due to its ability to generate remarkably complex behavior from a simple set of deterministic local rules. The model consists of a two-dimensional lattice in which each cell can assume one of two possible states, alive or dead. At each discrete time step, all cells are updated synchronously according to the states of their eight neighboring cells, known as the Moore neighborhood. Despite its simplicity, the repeated application of these local rules gives rise to a rich variety of emergent phenomena, including stable structures, oscillators, moving patterns (gliders), and highly complex interactions \cite{gameoflife1}. The name \emph{Life}\footnote{The game was named \emph{Life} because its rules mimic, in a highly simplified manner, fundamental biological processes such as survival, reproduction, and death under constraints imposed by local population density.} reflects the biological inspiration behind the model. The state transition of each cell is governed by the following rules:
\begin{itemize}
\item A live cell with fewer than two live neighbors dies (underpopulation).
\item A live cell with two or three live neighbors survives. (survival)
\item A live cell with more than three live neighbors dies (overpopulation).
\item A dead cell with exactly three live neighbors becomes alive (birth).
\end{itemize}

The general formulation of Definition~\ref{def:CA} encompasses a broad family of cellular automata. Conway's Game of Life is a particular instance of this framework, obtained by specifying the lattice, state space, neighborhood template, and local update rule as follows.

\begin{definition}[Conway's Game of Life]
Conway's Game of Life (GoL) is the two-dimensional cellular automaton
$\left\langle \mathbb Z^2, S, \Delta,\phi\right\rangle$,
where:
\begin{enumerate}
    \item The infinite lattice $\mathbb{Z}^2$;
    \item $S=\{0,1\}$ is the binary state set, where $0$ and $1$ denote dead and live cells, respectively;
    \item $\Delta$ is the set of Moore closed neighborhoods of range $1$, i.e., using a lexicographic order $x=(i,j)$:
    \begin{equation}
        \overline{\partial}_x=
        \big(
        (i,j),
        (i-1,j-1),(i-1,j),(i-1,j+1),
        (i,j-1),(i,j+1),
        (i+1,j-1),(i+1,j),(i+1,j+1)
        \big),
        \qquad m=\#\overline{\partial}_x=9.
    \end{equation}
    \item Denote 
    \begin{equation}
        C(x) = \sum c_{t}({\partial}_x)
    \end{equation}
    the number of living neighbors of cell $x$ (excluding itself). 
    The update rule is
    \begin{equation}
        \phi\big(c_t(\overline{\partial}_x)\big)=
        \begin{cases}
            1, & \text{if } c_t(x)=1 \text{ and } C(x)\in\{2,3\},\\
            1, & \text{if } c_t(x)=0 \text{ and } C(x)=3,\\
            0, & \text{otherwise}.
        \end{cases}
    \end{equation}
\end{enumerate}
\end{definition}

GoL's rule-set belongs to Wolfram Class IV \citep{Wolfram2002}, the class of rules that generate complex, localized structures capable of long-range interaction and information transport, rather than simply dying out (Class I/II) or degenerating into unstructured noise (Class III). 
Remarkably, it has been proven to be Turing complete \citep{rendell2011turing}, meaning that suitable initial configurations can implement arbitrary logical circuits and emulate the computation of any Turing machine. 
This computational universality has made the Game of Life a canonical benchmark for studying complexity, self-organization, and emergent computation \citep{WOLFRAM19841, rendell2011turing}. 
The book by
\citet{johnston2021conway} provides a comprehensive treatment of GoL's mathematical foundations, cataloging its spatial patterns, cycles, and periodic structures. 
\citet{ConwaysGameo2024l} propose quantitative tests for \textit{Life-like}\footnote{A \emph{Life-like} CA shares GoL's lattice, states, and Moore neighborhood but allows arbitrary birth/survival neighbor-count sets $B,S\subseteq\{0,\dots,8\}$ (B$/$S notation), yielding $2^{18}=262144$ possible rules; GoL is B3/S23.} behavior grounded in critical brain theory, Shannon's information entropy, and integrated information theory. 
More classical contributions include \citet{Speck1978}, who develop statistical mechanics approaches to GoL's large-scale dynamics and \citet{Conway19704} that introduced the rule and its specifications.

\subsection{Complexity and Entropy in Cellular Automata}

The investigation of dynamical complexity in cellular automata dates back to the late 1970s. \citet{Speck1978} studied the statistical mechanics of Conway's Game of Life and introduced a coarse-grained entropy measure to quantify spatial organization. The lattice was divided into smaller regions, and the number of live cells within each region was used to estimate the number of microscopic configurations compatible with the observed macroscopic state. Lower entropy values indicated greater spatial organization and stronger correlations, whereas higher values corresponded to more disordered configurations.

Later studies extended the analysis of cellular automata toward information-theoretic measures of disorder and organization. The maximum nonsymmetric entropy principle proposed by Liu \citep{Liu2008,Liu2009} was investigated in the context of Conway's Game of Life by \citet{ContrerasReyes2021Lerch}. Building on the connection between Conway's Game of Life and Shannon entropy investigated by \citet{pan2010emergencesymmetrynewtype}, the authors introduced an auxiliary information parameter that allows nonsymmetric entropy to be decomposed into Shannon entropy and negentropy. 
Applied to GoL, this framework was used to characterize the decay of the living-cell population over time and to obtain a global measure of disorder associated with the evolving population. 
The authors further suggested that the entropy--complexity plane could provide an information-based framework for studying and maximizing complexity in both discrete systems, such as cellular automata, and continuous systems.

In the context of information-theoretic analysis of cellular automata, \citet{Kharazmi2023deng} employed the Deng-Fisher information measure and its extensions to quantify uncertainty and characterize structural changes in Conway's Game of Life, demonstrating the potential of information-theoretic measures for analyzing the complex dynamics of cellular automata.

More recently, \citet{ConwaysGameo2024l} used entropy to characterize population-level disorder in cellular automata, treating a $50\%$ live-cell ``soup'' as maximally disordered. 
By comparing Conway's Game of Life with 100 randomly selected \textit{Life-like} cellular automata, the authors found that GoL exhibited unusually high sensitivity to initial conditions and evolved toward a relatively low-entropy population equilibrium. 
However, this entropy measure captures binary disorder in the live/dead-cell population rather than the full informational complexity or spatial organization of the configurations.

While entropy-based approaches quantify aspects of disorder and information, other studies have focused on characterizing the dynamical regimes generated by cellular automata. 
Langton formalized this perspective through the introduction of the $\lambda$ parameter, a statistical measure characterizing the dynamical regime of a cellular automaton according to the proportion of active transitions in its local update rule \citep{Langton1990Computation}. 
Considering a cellular automaton with $K$ possible states and a neighborhood of size $N$, the local rule contains $K^{K^N}$ possible rule tables. 
The $\lambda$ parameter is defined as the fraction of rule-table entries that produce non-quiescent states, $\lambda = {n_{\mathrm{active}}}/{K^{K^N}}$, where $n_{\mathrm{active}}$ denotes the number of rule-table entries that produce active states.
By varying $\lambda$, Langton demonstrated that cellular automata can exhibit distinct dynamical regimes. 
Low values of $\lambda$ generally correspond to ordered behavior, characterized by stable or periodic structures, whereas high values are associated with chaotic dynamics and increased sensitivity to initial conditions. 
Between these regimes lies a critical region, commonly referred to as the ``\textit{edge of chaos}'' \citep{Mitchell1993}, where ordered structures coexist with irregular dynamics and complex spatiotemporal behavior can emerge.

These studies substantiate that entropy, information, and rule-based parameters provide complementary perspectives on cellular automata dynamics. 
Entropy-based measures primarily characterize disorder, while parameters such as $\lambda$ characterize the dynamical regime imposed by the transition rule. 
This distinction motivates the use of measures that jointly capture disorder and structured organization, such as the entropy--complexity ($H\times C$) plane, which provide a framework for analyzing cellular automata trajectories in terms of both their degree of unpredictability and their statistical complexity.

In the following, we explain how we convert a CA generation into a time series for subsequent transformation into a sequence of values and then patterns.

\section{Peano-Hilbert Curve}

The Peano-Hilbert curve, a variant of the broader class of space-filling curves \citep{kanungo2024spacefillingcurves}, is a linearization method for multidimensional grids, e.g., an image or a CA lattice, through a 1D mapping. 
Its discrete, recursive formulation has since been widely adopted for indexing multidimensional data and preserving spatial correlations during dimensionality reduction \cite{Bader2013, Moon2001}. 
It provides a locality-preserving traversal of the $n$D lattice, in which spatially close cells tend to remain close in the resulting one-dimensional ordering.
Hilbert curves have previously been integrated into CA-based frameworks by \cite{Macedo2017Improving} in the context of automatically generating maps for game levels through procedural content generation. 
In that work, the main purpose of applying the Peano-Hilbert curve was to provide a global structural skeleton that preserves spatial connectivity alongside local automaton dynamics.

\begin{figure}[]
\centering
\begin{subfigure}[b]{0.3\textwidth}
\centering
\begin{tikzpicture}[scale=0.5]
\draw[step=1cm,gray!40] (0,0) grid (4,4);
\draw[thick,->]
(0.5,0.5)
--(1.5,0.5)
--(1.5,1.5)
--(0.5,1.5)
--(0.5,2.5)
--(0.5,3.5)
--(1.5,3.5)
--(1.5,2.5)
--(2.5,2.5)
--(2.5,3.5)
--(3.5,3.5)
--(3.5,2.5)
--(3.5,1.5)
--(2.5,1.5)
--(2.5,0.5)
--(3.5,0.5)
;
\node at (2,-0.5) {\footnotesize $K=2$ ($4\times4$)};
\end{tikzpicture}
\caption{}
\label{fig:hilbert_4x4}
\end{subfigure}
\hfill
\begin{subfigure}[b]{0.3\textwidth}
\centering
\begin{tikzpicture}[scale=0.5]
\draw[step=1cm,gray!40] (0,0) grid (8,8);
\draw[thick,->]
(0.5,0.5)
--(0.5,1.5)
--(1.5,1.5)
--(1.5,0.5)
--(2.5,0.5)
--(3.5,0.5)
--(3.5,1.5)
--(2.5,1.5)
--(2.5,2.5)
--(3.5,2.5)
--(3.5,3.5)
--(2.5,3.5)
--(1.5,3.5)
--(1.5,2.5)
--(0.5,2.5)
--(0.5,3.5)
--(0.5,4.5)
--(1.5,4.5)
--(1.5,5.5)
--(0.5,5.5)
--(0.5,6.5)
--(0.5,7.5)
--(1.5,7.5)
--(1.5,6.5)
--(2.5,6.5)
--(2.5,7.5)
--(3.5,7.5)
--(3.5,6.5)
--(3.5,5.5)
--(2.5,5.5)
--(2.5,4.5)
--(3.5,4.5)
--(4.5,4.5)
--(5.5,4.5)
--(5.5,5.5)
--(4.5,5.5)
--(4.5,6.5)
--(4.5,7.5)
--(5.5,7.5)
--(5.5,6.5)
--(6.5,6.5)
--(6.5,7.5)
--(7.5,7.5)
--(7.5,6.5)
--(7.5,5.5)
--(6.5,5.5)
--(6.5,4.5)
--(7.5,4.5)
--(7.5,3.5)
--(7.5,2.5)
--(6.5,2.5)
--(6.5,3.5)
--(5.5,3.5)
--(4.5,3.5)
--(4.5,2.5)
--(5.5,2.5)
--(5.5,1.5)
--(4.5,1.5)
--(4.5,0.5)
--(5.5,0.5)
--(6.5,0.5)
--(6.5,1.5)
--(7.5,1.5)
--(7.5,0.5)
;
\node at (4,-0.5) {\footnotesize $K=3$ ($8\times8$)};
\end{tikzpicture}
\caption{}
\label{fig:hilbert_8x8}
\end{subfigure}
\hfill
\begin{subfigure}[b]{0.3\textwidth}
\centering
\begin{tikzpicture}[scale=0.25]
\draw[step=1cm,gray!40] (0,0) grid (16,16);
\draw[thick,->]
(0.5,0.5)
--(1.5,0.5)
--(1.5,1.5)
--(0.5,1.5)
--(0.5,2.5)
--(0.5,3.5)
--(1.5,3.5)
--(1.5,2.5)
--(2.5,2.5)
--(2.5,3.5)
--(3.5,3.5)
--(3.5,2.5)
--(3.5,1.5)
--(2.5,1.5)
--(2.5,0.5)
--(3.5,0.5)
--(4.5,0.5)
--(4.5,1.5)
--(5.5,1.5)
--(5.5,0.5)
--(6.5,0.5)
--(7.5,0.5)
--(7.5,1.5)
--(6.5,1.5)
--(6.5,2.5)
--(7.5,2.5)
--(7.5,3.5)
--(6.5,3.5)
--(5.5,3.5)
--(5.5,2.5)
--(4.5,2.5)
--(4.5,3.5)
--(4.5,4.5)
--(4.5,5.5)
--(5.5,5.5)
--(5.5,4.5)
--(6.5,4.5)
--(7.5,4.5)
--(7.5,5.5)
--(6.5,5.5)
--(6.5,6.5)
--(7.5,6.5)
--(7.5,7.5)
--(6.5,7.5)
--(5.5,7.5)
--(5.5,6.5)
--(4.5,6.5)
--(4.5,7.5)
--(3.5,7.5)
--(2.5,7.5)
--(2.5,6.5)
--(3.5,6.5)
--(3.5,5.5)
--(3.5,4.5)
--(2.5,4.5)
--(2.5,5.5)
--(1.5,5.5)
--(1.5,4.5)
--(0.5,4.5)
--(0.5,5.5)
--(0.5,6.5)
--(1.5,6.5)
--(1.5,7.5)
--(0.5,7.5)
--(0.5,8.5)
--(0.5,9.5)
--(1.5,9.5)
--(1.5,8.5)
--(2.5,8.5)
--(3.5,8.5)
--(3.5,9.5)
--(2.5,9.5)
--(2.5,10.5)
--(3.5,10.5)
--(3.5,11.5)
--(2.5,11.5)
--(1.5,11.5)
--(1.5,10.5)
--(0.5,10.5)
--(0.5,11.5)
--(0.5,12.5)
--(1.5,12.5)
--(1.5,13.5)
--(0.5,13.5)
--(0.5,14.5)
--(0.5,15.5)
--(1.5,15.5)
--(1.5,14.5)
--(2.5,14.5)
--(2.5,15.5)
--(3.5,15.5)
--(3.5,14.5)
--(3.5,13.5)
--(2.5,13.5)
--(2.5,12.5)
--(3.5,12.5)
--(4.5,12.5)
--(5.5,12.5)
--(5.5,13.5)
--(4.5,13.5)
--(4.5,14.5)
--(4.5,15.5)
--(5.5,15.5)
--(5.5,14.5)
--(6.5,14.5)
--(6.5,15.5)
--(7.5,15.5)
--(7.5,14.5)
--(7.5,13.5)
--(6.5,13.5)
--(6.5,12.5)
--(7.5,12.5)
--(7.5,11.5)
--(7.5,10.5)
--(6.5,10.5)
--(6.5,11.5)
--(5.5,11.5)
--(4.5,11.5)
--(4.5,10.5)
--(5.5,10.5)
--(5.5,9.5)
--(4.5,9.5)
--(4.5,8.5)
--(5.5,8.5)
--(6.5,8.5)
--(6.5,9.5)
--(7.5,9.5)
--(7.5,8.5)
--(8.5,8.5)
--(8.5,9.5)
--(9.5,9.5)
--(9.5,8.5)
--(10.5,8.5)
--(11.5,8.5)
--(11.5,9.5)
--(10.5,9.5)
--(10.5,10.5)
--(11.5,10.5)
--(11.5,11.5)
--(10.5,11.5)
--(9.5,11.5)
--(9.5,10.5)
--(8.5,10.5)
--(8.5,11.5)
--(8.5,12.5)
--(9.5,12.5)
--(9.5,13.5)
--(8.5,13.5)
--(8.5,14.5)
--(8.5,15.5)
--(9.5,15.5)
--(9.5,14.5)
--(10.5,14.5)
--(10.5,15.5)
--(11.5,15.5)
--(11.5,14.5)
--(11.5,13.5)
--(10.5,13.5)
--(10.5,12.5)
--(11.5,12.5)
--(12.5,12.5)
--(13.5,12.5)
--(13.5,13.5)
--(12.5,13.5)
--(12.5,14.5)
--(12.5,15.5)
--(13.5,15.5)
--(13.5,14.5)
--(14.5,14.5)
--(14.5,15.5)
--(15.5,15.5)
--(15.5,14.5)
--(15.5,13.5)
--(14.5,13.5)
--(14.5,12.5)
--(15.5,12.5)
--(15.5,11.5)
--(15.5,10.5)
--(14.5,10.5)
--(14.5,11.5)
--(13.5,11.5)
--(12.5,11.5)
--(12.5,10.5)
--(13.5,10.5)
--(13.5,9.5)
--(12.5,9.5)
--(12.5,8.5)
--(13.5,8.5)
--(14.5,8.5)
--(14.5,9.5)
--(15.5,9.5)
--(15.5,8.5)
--(15.5,7.5)
--(14.5,7.5)
--(14.5,6.5)
--(15.5,6.5)
--(15.5,5.5)
--(15.5,4.5)
--(14.5,4.5)
--(14.5,5.5)
--(13.5,5.5)
--(13.5,4.5)
--(12.5,4.5)
--(12.5,5.5)
--(12.5,6.5)
--(13.5,6.5)
--(13.5,7.5)
--(12.5,7.5)
--(11.5,7.5)
--(11.5,6.5)
--(10.5,6.5)
--(10.5,7.5)
--(9.5,7.5)
--(8.5,7.5)
--(8.5,6.5)
--(9.5,6.5)
--(9.5,5.5)
--(8.5,5.5)
--(8.5,4.5)
--(9.5,4.5)
--(10.5,4.5)
--(10.5,5.5)
--(11.5,5.5)
--(11.5,4.5)
--(11.5,3.5)
--(11.5,2.5)
--(10.5,2.5)
--(10.5,3.5)
--(9.5,3.5)
--(8.5,3.5)
--(8.5,2.5)
--(9.5,2.5)
--(9.5,1.5)
--(8.5,1.5)
--(8.5,0.5)
--(9.5,0.5)
--(10.5,0.5)
--(10.5,1.5)
--(11.5,1.5)
--(11.5,0.5)
--(12.5,0.5)
--(13.5,0.5)
--(13.5,1.5)
--(12.5,1.5)
--(12.5,2.5)
--(12.5,3.5)
--(13.5,3.5)
--(13.5,2.5)
--(14.5,2.5)
--(14.5,3.5)
--(15.5,3.5)
--(15.5,2.5)
--(15.5,1.5)
--(14.5,1.5)
--(14.5,0.5)
--(15.5,0.5)
;
\node at (8,-1) {\footnotesize $K=4$ ($16\times16$)};
\end{tikzpicture}
\caption{}
\label{fig:hilbert_16x16}
\end{subfigure}
\caption{Discrete Hilbert curves for orders $K=2$, $K=3$, and $K=4$, on meshes $4\times4$, $8\times8$, and $16\times16$, respectively. The traversal defines a bijective mapping between two-dimensional spatial coordinates and a one-dimensional ordering.}
\label{fig:hilbert_curves}
\end{figure}
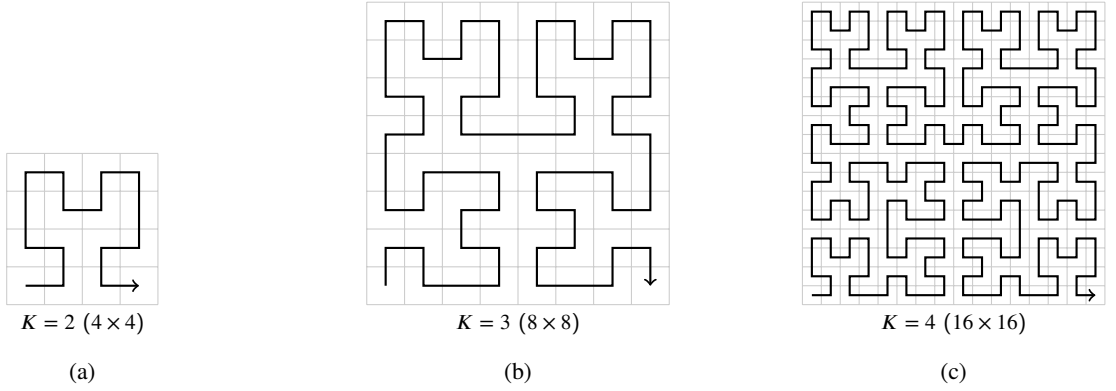

In the following, we will consider dynamics defined on lattices of dimension $2^K\times 2^K$.
Figure~\ref{fig:hilbert_curves} illustrates these lattices for $4\times4$, $8\times8$, and $16\times16$, showing how the curve's trajectory scans the grid while preserving spatial structure.

The Hilbert-Peano linearization consists of representing
the generation $c_t$ as a sequence $c_t(x)$ such that $x$ obeys the rule depicted in Figure~\ref{fig:hilbert_curves}.
We denote the resulting sequence of cell states, at generation $t$ as
\begin{equation}
    X_t = \big(s_1, s_2, \ldots, s_{2^{2K}}\big).
\end{equation}
This representation can subsequently be treated as a time series, 
allowing methods originally developed for one-dimensional temporal data, 
such as ordinal-pattern analysis, 
to be applied to the spatial organization of the cellular automaton.

\section{Ordinal Patterns}
First introduced by \citet{bandt2002permutation}, the ordinal patterns methodology has been widely applied to the analysis and characterization of time series, providing insights into a broad range of dynamical behaviors, including chaotic dynamics, stochastic processes, white noise, and regular oscillations \citep{bandt2002permutation, Leyva_2022}. The underlying idea is to characterize the temporal dynamics of a system through the ordinal relationships among successive observations, thereby providing an information-theoretic representation of its temporal structure. By quantifying the distribution of ordinal patterns, this approach enables the characterization of the degree of disorder, regularity, and complexity exhibited by the observed dynamics \citep{chagas2022white, Leyva_2022, Giri2025Permutation}.

\subsection{Definition}
Consider a time series $X_t = (x_1, \dots, x_\ell)$. The ordinal pattern methodology maps subsequences of $X_t$ into symbols that represent the relative ordering of their values rather than the values themselves. This transformation is invariant under nonlinear monotonic transformations of the data and requires no prior assumptions about the underlying probability distribution.

Given an embedding dimension $D \geq 2$ and a time delay $\tau \geq 1$ (typically $\tau = 1$), each point $x_t$ is associated with a $D$-dimensional vector
\begin{equation}
\mathbf{s}_t = (x_t, x_{t+\tau}, x_{t+2\tau}, \dots, x_{t+(D-1)\tau}).
\end{equation}

For each vector $\mathbf{s}_t$, the indices $(0, 1, \dots, D-1)$ are reordered according to the ascending order of the corresponding values, yielding a permutation
\begin{equation}
\pi = (r_0, r_1, \dots, r_{D-1})
\end{equation}
such that $x_{t+r_0\tau} \leq x_{t+r_1\tau} \leq \dots \leq x_{t+r_{D-1}\tau}$. 
Ties are broken according to the index order, following the convention adopted by \citet{bandt2002permutation}. 
Each $\mathbf{s}_t$ is therefore associated with one of the $D!$ possible ordinal patterns (permutations) of length $D$.

Sliding this window across the series produces a sequence of ordinal patterns, from which the relative frequency of each of the $D!$ possible permutations $\pi_i$ is estimated as
\begin{equation}
p(\pi_i) =
\frac{\text{number of occurrences of } \pi_i}
{\ell-(D-1)\tau},
\qquad i=1,\ldots,D!.
\end{equation}

This defines an empirical probability distribution $P = \big(p(\pi_1), \dots, p(\pi_{D!})\big)$ over the space of ordinal patterns, which characterizes the ordering structure of the time series independently of its absolute amplitude values.

In the following, we define the two most used features extracted from $P$, namely, 
the permutation entropy
and the statistical complexity.

\subsection{Permutation Entropy}
Given $P$, the normalized Shannon permutation entropy is defined as
\begin{equation}
H[P] = -\frac{1}{\log(D!)} \sum_{i=1}^{D!} p(\pi_i) \log p(\pi_i), \qquad H[P] \in [0,1],
\end{equation}
with the convention $0\log(0)=0$.
This quantity is also known as permutation entropy.
Distributions concentrated on a single ordinal pattern produce
$H[P] = 0$, indicating a highly regular ordering structure, whereas $H[P] = 1$ corresponds to a uniform distribution over all possible ordinal patterns, indicating the highest degree of ordinal diversity.

\subsection{Statistical Complexity}

While permutation entropy quantifies the degree of disorder or uncertainty in the ordinal-pattern distribution, it does not distinguish between distributions with different degrees of structural organization. 
To complement the entropy measure, we employ the statistical complexity measure based on the Jensen-Shannon divergence \citep{Rosso2007}.
    
Let $P_e$ denote the uniform distribution on $D!$ states:
    \begin{equation}
    P_e = \Big(\frac{1}{D!},\dots,\frac{1}{D!}\Big).
    \end{equation}
    
The Jensen-Shannon divergence between $P$ and $P_e$ is defined as    
    \begin{equation}
    J[P,P_e] =
    S\left[\frac{P+P_e}{2}\right]
    -\frac{1}{2}S[P]
    -\frac{1}{2}S[P_e],
    \end{equation}
    where $S[P]$ is the Shannon entropy of the distribution $P$,    
    \begin{equation}
    S[P] = -\sum_{i=1}^{D!}p(\pi_i)\log p(\pi_i).
    \end{equation}
    
    The statistical complexity is then obtained by normalizing the Jensen--Shannon divergence and weighting it by the normalized permutation entropy,   
    \begin{equation}
    C[P] = Q[P,P_e],H[P],
    \end{equation}
    where
    \begin{equation}
    Q[P,P_e] = \frac{J[P,P_e]}{J_{\max}}
    \end{equation}
    is the normalized Jensen-Shannon divergence, and $J_{\max}$ denotes its maximum value for a distribution with $D!$ possible ordinal patterns.
The resulting statistical complexity satisfies $C[P]\in[0,1]$. 

In contrast to entropy alone, the complexity measure quantifies the degree of structure or organization present in the ordinal-pattern distribution. 
In particular, distributions concentrated on a single ordinal pattern have low entropy and low complexity, while the uniform distribution has maximum entropy but vanishing complexity. 
Intermediate values of entropy and complexity characterize distributions that exhibit both disorder and non-trivial structure.

\subsection{The entropy--complexity plane} 

The entropy--complexity plane ($H\times C$) has the ability to represent the dynamical properties of time series by combining normalized permutation entropy and statistical complexity. 
Each time series is represented by a point
    \begin{equation}
    (H[P],C[P]),
    \end{equation}    
    where $H[P]$ measures the diversity of ordinal patterns and $C[P]$ quantifies the degree of non-trivial structure in their probability distribution.
    The plane is a closed manifold embedded in $[0,1]^2$.
    This representation makes it possible to compare different dynamical regimes according to their degree of disorder and structural organization. 
    
    In the context of the cellular automaton considered in this work, the trajectory of the system in the $H\times C$ plane can be used to characterize changes in the information structure of its spatiotemporal dynamics.

\section{Proposed Method} \label{sec:proposedmethod}

The main goal of this work is to propose and apply a method for mapping the dynamics of the Game of Life cellular automaton onto the entropy--complexity plane, thereby providing a representation of the information structure underlying its spatiotemporal behavior. 
The following sections describe the proposed methodology and present the analyses derived from this representation.

\subsection{$H\times C$ Trajectory}

Consider a finite two-dimensional square lattice of side $2^K$, 
an instance of Definition~\ref{def:CA} with state set $S=\{0,1\}$ (dead/alive), 
Moore neighborhood $( m=\#\overline{\partial}_x=9)$, and periodic (toroidal) boundary conditions. 
The Game of Life (GoL) local update rule $\phi$ is Eq.~\eqref{eq:TransitionFunction}. 
Given $0<\rho_0<100$, each cell is set alive or dead independently to obtain $c_0$, the initial state with exactly $\rho\%$ of live cells.
Given a finite simulation horizon $T$, a generation at time $t\in 1,\dots,T$ denotes the configuration $c_{t}$ obtained after $t-1$ applications of $\phi$.

The proposed methodology proceeds as follows:
\begin{enumerate}
    \item\label{Step1} At each generation $t$, the lattice is evolved by one GoL time step per Eq.~\eqref{eq:TransitionFunction}, producing $c_{t}$ from $c_{t-1}$.

    \item The $2^K\times 2^K$ lattice $c_{t}$ is traversed using a discrete Peano-Hilbert space-filling curve of order $K$.

    \item The binary values of $c_{t}$ are grouped into non-overlapping blocks of $b$ consecutive cells (bits), each converted to its integer value in $\{0,\dots,2^b-1\}$,
yielding a compact scalar representation of $b$ spatially-local cells. 

    \item The resulting values are a time series $X_t=(x_1,\dots,x_\ell)$ of length $\ell=2^{2K}/b$, describing the spatial configuration at generation $t$. This time series is the input for the ordinal-pattern framework with embedding dimension $D$, yielding the permutation probability distribution $\hat P_t=(\hat p_1,\dots,\hat p_{D!})$.

    \item\label{Step5} The normalized Shannon entropy $H[\hat P_t]$ and the Jensen-Shannon statistical complexity $C[\hat P_t]$ are computed from $\hat P_t$.

    \item Steps~\ref{Step1} to \ref{Step5} are repeated for every generation $t\in 1,\dots,T $, producing a trajectory $\{(H_t,C_t)\}_{t=1}^{T}$, with $H_t\equiv H[\hat P_t]$ and $C_t\equiv C[\hat P_t]$, on the $H\times C$ plane characterizing the automaton's dynamical complexity over time.
\end{enumerate}

\begin{figure*}[]
    \centering
    \includegraphics[width=0.8\linewidth]{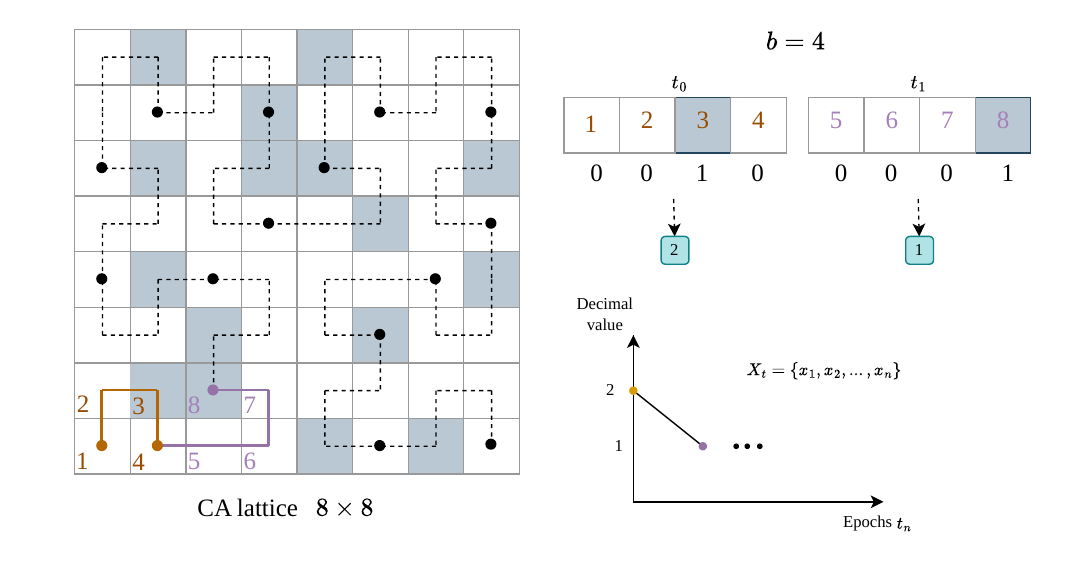}
    \caption{Example of the time-series construction from a $8\times 8 $ CA lattice with $b=4$.}
    \label{fig:met}
\end{figure*}

Figure~\ref{fig:met} illustrates the construction of the time series from a single generation $t$ of an $8\times8$ CA lattice with $b=4$. As detailed in Algorithm~\ref{alg:hxc_gol}, this procedure is repeated for every generation of the CA, resulting in a historical time series that encodes the informational structure of the system's spatiotemporal dynamics.

\begin{algorithm}
\caption{Hilbert Curve-Based entropy--complexity Analysis of Conway's Game of Life}
\label{alg:hxc_gol}
\begin{algorithmic}[1]

\Require initial configuration $c_{0}$ of size $2^K\times 2^K$, number of generations $T$,
Hilbert order $K$, block size $b$, embedding dimension $D$
\Ensure trajectory $\{(H_t,C_t)\}_{t=1}^{T}$ on the entropy--complexity plane

\Statex
\Function{BuildHilbertMap}{$K$}
\Comment{Construct the discrete Peano-Hilbert ordering}
\State $N \gets 2^{2K}$
\Comment{Total number of lattice cells}
\State $h_K \gets$ bijection from $\mathcal{L}_2$ to $\{0,1,\dots,N-1\}$
\State \Return $h_K$, $N$
\EndFunction

\Statex
\Function{ConfigurationToSequence}{$c_{(t)}$, $h_K$, $N$, $b$}
\Comment{Convert the 2D configuration into scalar symbols}
\State $\ell \gets N/b$
\State $X_t \gets (x_1,\dots,x_\ell)$
\For{$j=1$ to $\ell$}
\State $x_j \gets 0$
\For{$k=1$ to $b$}
\State $x_j \gets x_j +
c_{(t)}\big(h_K^{-1}((j-1)b+k-1)\big)\cdot 2^{b-k}$
\EndFor
\EndFor
\State \Return $X_t$
\EndFunction

\Statex
\Function{EstimateEntropyComplexity}{$X_t$, $D$}
\Comment{Estimate ordinal-pattern entropy and statistical complexity}
\State $\hat P_t \gets (\hat p_1,\dots,\hat p_{D!})$
\Comment{Permutation probability distribution of $X_t$}
\State $H_t \gets H[\hat P_t]$
\Comment{Normalized Shannon entropy}
\State $C_t \gets C[\hat P_t]$
\Comment{Jensen-Shannon statistical complexity}
\State \Return $(H_t,C_t)$
\EndFunction

\Statex
\Procedure{HxCTrajectory}{$c_{(0)}$, $T$, $K$, $b$, $D$}
\State $(h_K, N) \gets$ \Call{BuildHilbertMap}{$K$}
\State trajectory $\gets$ empty list
\State $c_{(t)} \gets c_{(0)}$
\For{$t=1$ to $T$}
\State $c_{(t)} \gets F(c_{(t-1)})$
\Comment{One GoL transition per Eq.~\eqref{eq:TransitionFunction}}
\State $X_t \gets$ \Call{ConfigurationToSequence}{$c_{(t)}$, $h_K$, $N$, $b$}
\State $(H_t,C_t) \gets$ \Call{EstimateEntropyComplexity}{$X_t$, $D$}
\State append $(H_t,C_t)$ to trajectory
\EndFor
\State \Return trajectory
\EndProcedure

\end{algorithmic}
\end{algorithm}

\subsection{Convergence to the Regime}
\label{sec:convergence}

To characterize how rapidly the automaton's trajectory in the $H \times C$ plane approaches its steady state behavior, we define a distance-to-limit and estimate its rate of decay.

Let $(H^*, C^*)$ denote the steady state of the trajectory, estimated as the sample mean of $(H_t, C_t)$ over the final $1<\upsilon<T$ generations of the simulation:
\begin{equation}\label{eq:h*c*mean}
    H^* = \frac{1}{\upsilon} \sum_{t=T-\upsilon+1}^{T} H_t, \qquad
    C^* = \frac{1}{\upsilon} \sum_{t=T-\upsilon+1}^{T} C_t.
\end{equation}

For each generation $t$, the Euclidean distance to this steady state point is computed as
\begin{equation}\label{eq:dt}
    d_t = \sqrt{(H_t - H^*)^2 + (C_t - C^*)^2}.
\end{equation}

An exponential relaxation model was chosen as it represents the generic first-order approximation of a discrete dynamical system's behavior in the neighborhood of an attracting state. 
Under this assumption, $d_t = d_0 e^{-rt}$, and the logarithmic transform $\log(d_t)$ reduces the parameter estimation to ordinary linear regression, with the convergence rate $r$ recovered directly from the fitted slope:
\begin{equation}\label{eq:log(dt)}
    \log(d_t) = \log(d_0) - r\,t + \varepsilon_t.
\end{equation}
The corresponding half-life, $t_{1/2} = \ln(2) / r$, represents the number of generations required for the distance to the steady state point to decrease by $50\%$ under the fitted exponential relaxation model. 
Thus, a smaller $t_{1/2}$ indicates faster convergence, whereas a larger value indicates slower convergence.

\subsection{Time Series Length $\ell$}
\label{sec:time_series_length}

It is established that the length of a time series should satisfy the condition $\ell \gg D!$ to obtain reliable statistics when estimating permutation-based quantities, since $D!$ represents the number of possible ordinal patterns for an embedding dimension $D$ \citep{staniek2007parameter,PallaresDiNunzio2025}. 
However, satisfying this condition alone does not guarantee reliable estimations. 
Even when the time series is sufficiently long to accommodate the $D!$ possible patterns, finite-sample effects can still introduce substantial sampling fluctuations in the estimated permutation entropy (PE) \citep{QuinteroQuiroz2015}.

In the proposed methodology, the time-series length is not an independent parameter, but is determined by the spatial resolution of the cellular automaton and the bit-depth used in the Peano--Hilbert encoding. 
Specifically, for a square lattice of size $2^K \times 2^K$ and a bit-depth $b$, the resulting time-series length is given by
\begin{equation} \label{eq:l=l2/b}
\ell = \frac{2^{2K}}{b}.
\end{equation}
Consequently, increasing the bit-depth reduces the number of observations available for the estimation of permutation entropy and statistical complexity, while simultaneously increasing the number of possible states in the encoded time series, from $2^b$ possible values (Table~\ref{tab:ca_timeseries_size_1024}). 
Notably, however, $b$ does not act merely as a parameter controlling the time-series length $\ell$: it also determines the granularity of the symbolic representation from which the time series is constructed. 
Specifically, different bit-depths aggregate different numbers of consecutive lattice cells along the Peano--Hilbert traversal into a single scalar value, thereby altering both the number of available observations and the state space of the resulting time series.

This dual role of the bit-depth makes the relationship between $\ell$ and the statistical reliability of the entropy--complexity estimates non-trivial. Although the ordinal-pattern analysis is formally determined by the embedding dimension $D$, rather than by the size of the encoded state space, changes in $b$ modify the statistical properties of the resulting time series and may consequently affect the estimation of $(H,C)$. Therefore, an empirical assessment is required to determine how the choice of $L$ and $b$ affects the resulting entropy--complexity representation and its variability across independent realizations.

\begin{table}[]
\caption{Time series size for a $1024\times1024$ cellular automaton lattice as a function of bit-depth $b$, computed as $N = 1024^2 / b$, alongside the number of possible states per cell, $2^b$.}\label{tab:ca_timeseries_size_1024}
\begin{tabular*}{\tblwidth}{@{}LLL@{}}
\toprule
$b$ (bits) & Time Series Size $\ell$ & $|\mathcal{S}|$ ($2^b$) \\ 
\midrule
8  & 131072 & $2.56\times10^{2}$ \\
16 & 65536  & $6.55\times10^{4}$ \\
32 & 32768  & $4.29\times10^{9}$ \\
\bottomrule
\end{tabular*}
\end{table}

\subsubsection{Sensitivity Analysis}
\label{sec:sensitivity_analysis}

To investigate the influence of bit-depth and lattice size on the resulting entropy--complexity representation, we experimentally evaluate the variability of the converged $(H,C)$ estimates obtained from different combinations of these parameters. 
For each configuration $(L,b)$, with the initial density of live cells fixed at $\rho_0=50\%$ and the embedding dimension set to $D=6$, $n_{\mathrm{rep}}=30$ independent Game of Life realizations are generated, with $10000$ generations each. 
The lattice size is varied as $K \in \{7,8,9,10,11\}$ and the bit-depth as $b \in \{8,16,32\}$.

For each realization $t$, the reference point $(H^*_{t,L,b},C^*_{t,L,b})$ is estimated according to Equation~\ref{eq:h*c*mean}. 
Specifically, the reference point is obtained by averaging the entropy and statistical complexity over the final $60\%$ of the simulated trajectory, with $\upsilon=0.4T$. 
This procedure provides a single representative point in the entropy--complexity plane for each realization and parameter configuration.

For each configuration $(L,b)$, the mean reference point across the $n_{\mathrm{rep}}=30$ independent realizations is computed as
\begin{equation}
\overline{H^*}_{L,b}
=
\frac{1}{n_{\mathrm{rep}}}
\sum_{t=1}^{n_{\mathrm{rep}}}
H^*_{t,L,b},
\qquad
\overline{C^*}_{L,b}
=
\frac{1}{n_{\mathrm{rep}}}
\sum_{t=1}^{n_{\mathrm{rep}}}
C^*_{t,L,b},
\end{equation}
with the corresponding standard deviations $\sigma_{H^*}$ and $\sigma_{C^*}$ computed across the independent realizations.

Because the bit-depth determines the number of states encoded in each element of the resulting time series, while the lattice size controls the total number of encoded elements, their combined effect may influence both the location and the variability of the resulting entropy--complexity point. 
For each configuration, the corresponding time-series length is given by Equation~\eqref{eq:l=l2/b}.

The resulting means and standard deviations are used to assess two complementary aspects of the parameter choice. First, the variation of $(\overline{H^*}_{L,b},\overline{C^*}_{L,b})$ across different values of $L$ for a fixed bit-depth provides an empirical measure of the sensitivity of the entropy--complexity representation to the resulting time-series length. Second, the standard deviations $\sigma_{H^*}$ and $\sigma_{C^*}$ across the 30 independent realizations quantify the robustness of the estimated reference point to differences in the initial conditions.

The analysis therefore evaluates whether increasing the lattice size produces substantial changes in the estimated reference point and whether the resulting estimates remain consistent across independent realizations. In parallel, comparisons across bit-depths reveal the extent to which the granularity of the spatial encoding affects the entropy--complexity representation independently of the resulting time-series length. These observations provide an empirical basis for selecting the bit-depth and lattice size used in the subsequent experiments, ensuring that the chosen configuration provides a sufficiently stable and reproducible representation of the long-run behavior of the cellular automaton.

\section{Experimental Design}

The parameter settings can be divided into three main categories: 
(i)~the CA simulation, defined by the number of generations and the initial density of living cells; 
(ii)~the ordinal-pattern analysis, defined by the embedding dimension; 
and the (iii)~Peano-Hilbert encoding, defined by the block size (in bits) used for the decimal conversion. 

The Game of Life (GoL) simulations were performed on a square lattice of ($2^{10} \times 2^{10}$) cells ($1024 \times 1024$) with toroidal boundary conditions. 
The initial configuration ($\text{generation}=1$) is the \textit{soup} strategy: 
cell states are randomly initialized to comply with the specified proportion of live cells ($\rho_0$). 
The simulation was carried over $10000$ generations. 
We used embedding dimensions $5$ and $6$, while the number of bits sampled along the Peano-Hilbert trajectory varied from $2^3$ to $2^{5}$. 
The complete grid of experimental parameters is presented in Table~\ref{tab:parameters}.

\begin{table}[]
\caption{Parameter values explored in the experiments.}\label{tab:parameters}
\begin{tabular*}{\tblwidth}{@{}LLL@{}}
\toprule
Description & Parameter & Values \\ %
\midrule
\multirow{2}{*}{CA Simulation} & Number of generations [$k$] & $\{10\}$ \\
 & Initial living-cells ($\rho_0\%$) & $\{10, 30, 50, 80\}$ \\
Ordinal Patterns & Embedding dimension & $\{5, 6\}$ \\
Peano-Hibert & Bits & $\{8, 16, 32\}$ \\ 
\bottomrule
\end{tabular*}
\end{table}

Appendix~\ref{app:ParametricSpace} describes results on a broader parameter space and justifies the choice presented in Table~\ref{tab:parameters}.

In terms of implementation details, all experiments were conducted in the R programming language. The \texttt{StatOrdPattHxC} package~\citep{packageROrd} was used to compute the ordinal pattern probabilities and estimate the corresponding points in the ($H \times C$) plane, while the \texttt{ggplot2} package was employed to generate the figures. 

\section{Results}

\subsection{Time Series Length Analysis}

Our first experiment determines the optimal lattice size for $D=6$, which is roughly the maximum embedding dimension used in the literature.
Results for smaller embedding dimensions will follow.

\begin{figure*}[]
    \centering
    \includegraphics[width=0.8\linewidth]{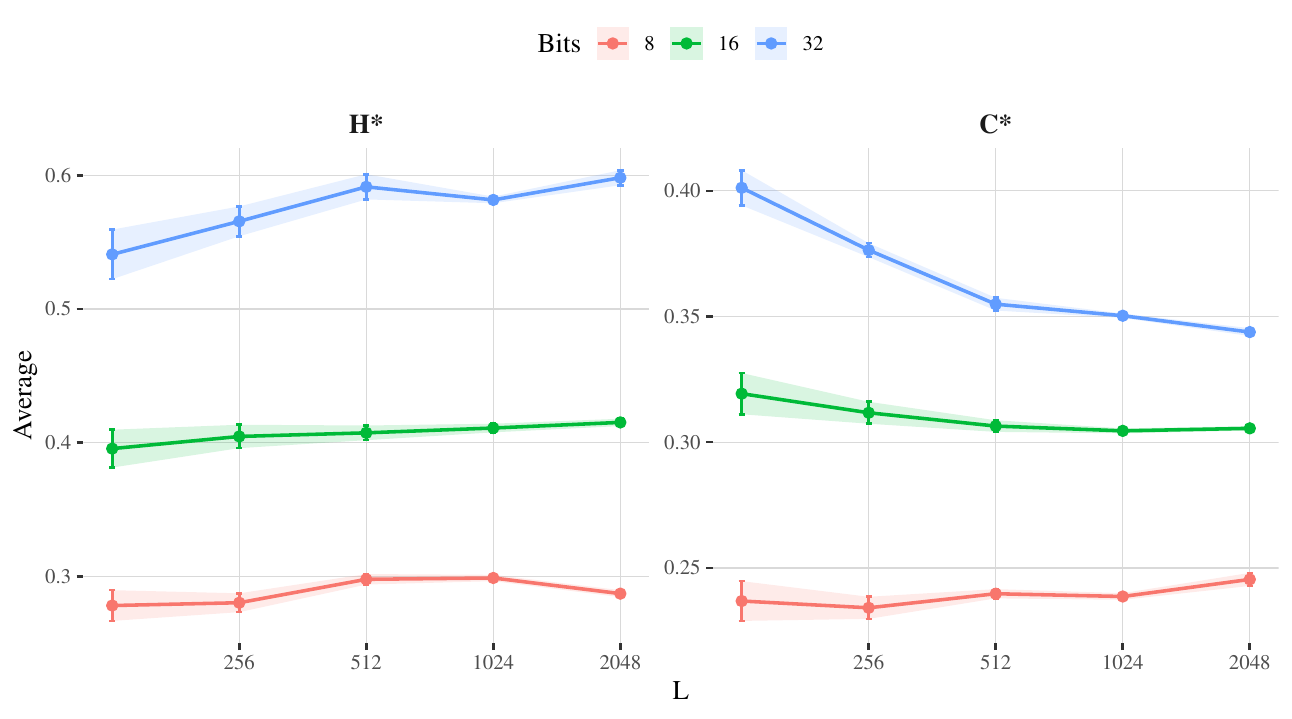}
    \caption{Sensitivity of the average permutation entropy $H^*$ (left) and complexity $C^*$ (right) to the lattice size $L$, for three bit-depths used to discrete the cellular automaton states ($b=8$, red; $b=16$, green; $b=32$, blue). Points denote the mean over three independent repetitions, error bars denote $\pm 1$ standard error, and shaded ribbons denote the corresponding confidence interval. }
    \label{fig:tslen}
\end{figure*}

Figure~\ref{fig:tslen} presents a sensitivity analysis 
of the observables at the steady state $(H^*, C^*)$ with respect to the bit-depth $b \in \{8, 16, 32\}$ and the lattice size $L \in \{128, 256, 512, 1024, 2048\}$. 
For each configuration, the resulting $(H^*, C^*)$ pair was averaged over $30$ independent repetitions, with shaded ribbons and error bars denoting the corresponding variability.

Bit-depth is the dominant factor governing the location of the trajectories in the entropy--complexity plane: the three curves occupy well-separated bands in both panels, with essentially no overlap even accounting for the confidence intervals.
Increasing $b$ from $8$ to $32$ bits shifts $H^*$ upward by roughly $0.30$ units (from $\approx 0.28$ to $\approx 0.58$) 
and shifts $C^*$ upward by a comparable margin at small $L$ (from $\approx 0.24$ to $\approx 0.40$ at $L=128$).
For this reason, our subsequent experiments analyze the dependence of the observable on the three discretization levels.

The main outcome from this analysis is that with $L = 1024$ it is possible to see a stability in the values of $(H^*,C^*)$ and therefore for the following study we adopt this CA lattice size.

\subsection{Entropy--Complexity Plane}

\begin{figure*}[]
\centering
\begin{subfigure}[b]{0.48\textwidth}
    \centering
    \includegraphics[width=\textwidth]{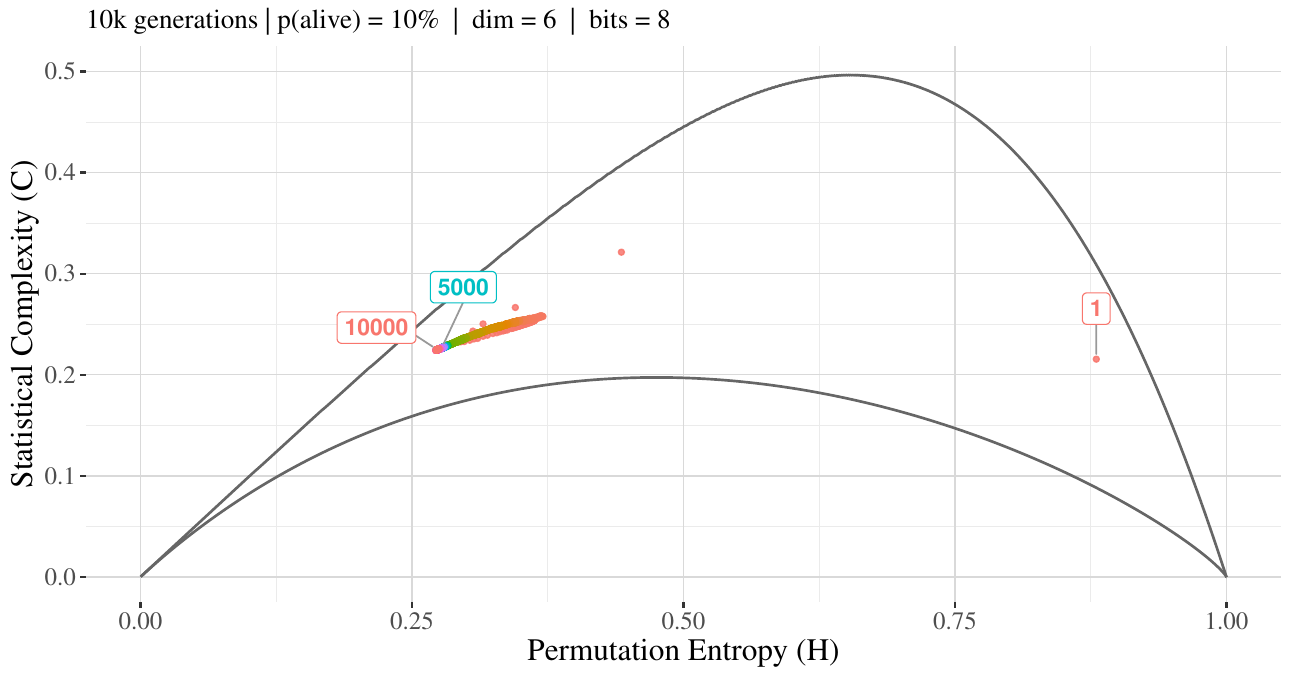}
    \caption{$\rho_0=10\%$}
    \label{fig:initial_10_8b}
\end{subfigure}
\hfill
\begin{subfigure}[b]{0.48\textwidth}
    \centering
    \includegraphics[width=\textwidth]{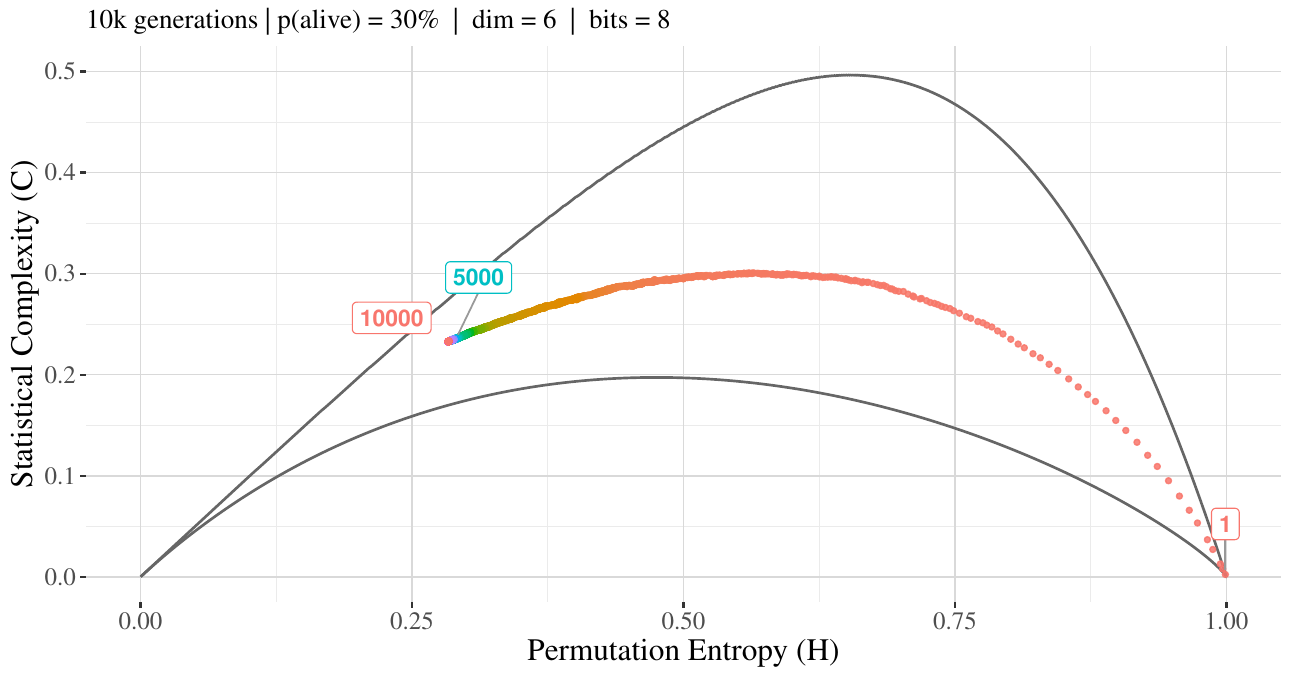}
    \caption{$\rho_0=30\%$}
    \label{fig:initial_30_8b}
\end{subfigure}
\vspace{0.5em}
\begin{subfigure}[b]{0.48\textwidth}
    \centering
    \includegraphics[width=\textwidth]{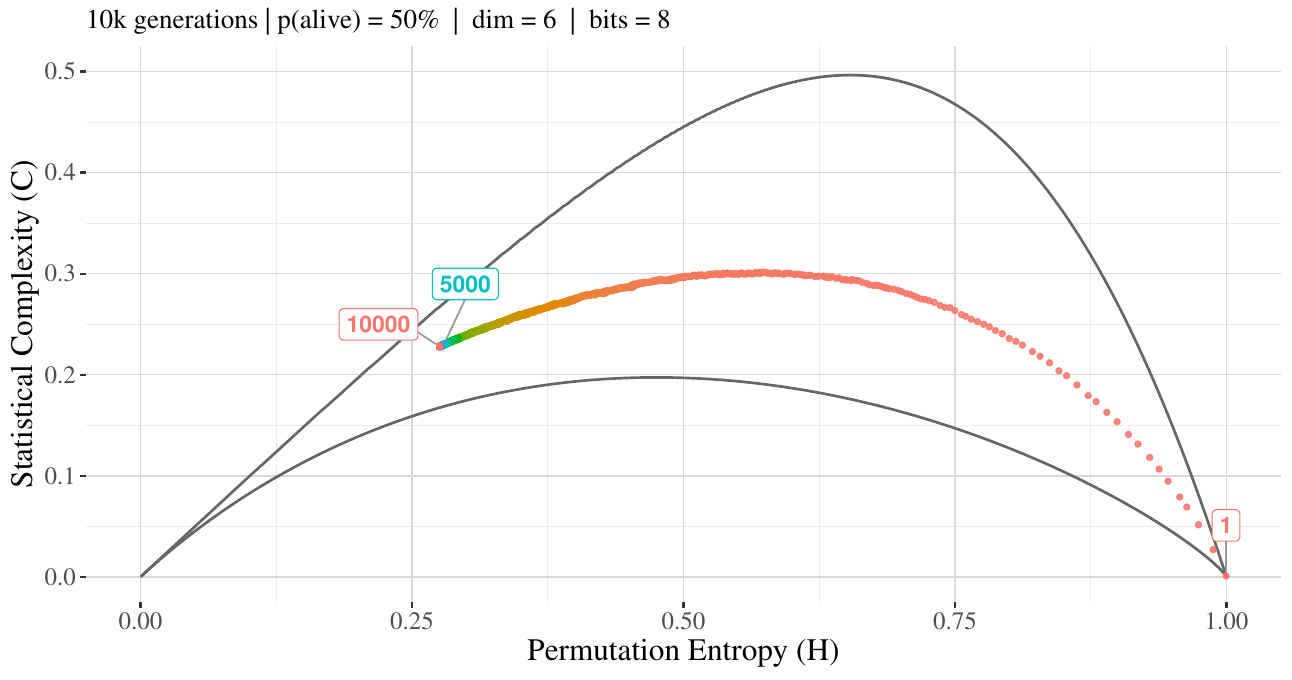}
    \caption{$\rho_0=50\%$}
    \label{fig:initial_50_8b}
\end{subfigure}
\hfill
\begin{subfigure}[b]{0.48\textwidth}
    \centering
    \includegraphics[width=\textwidth]{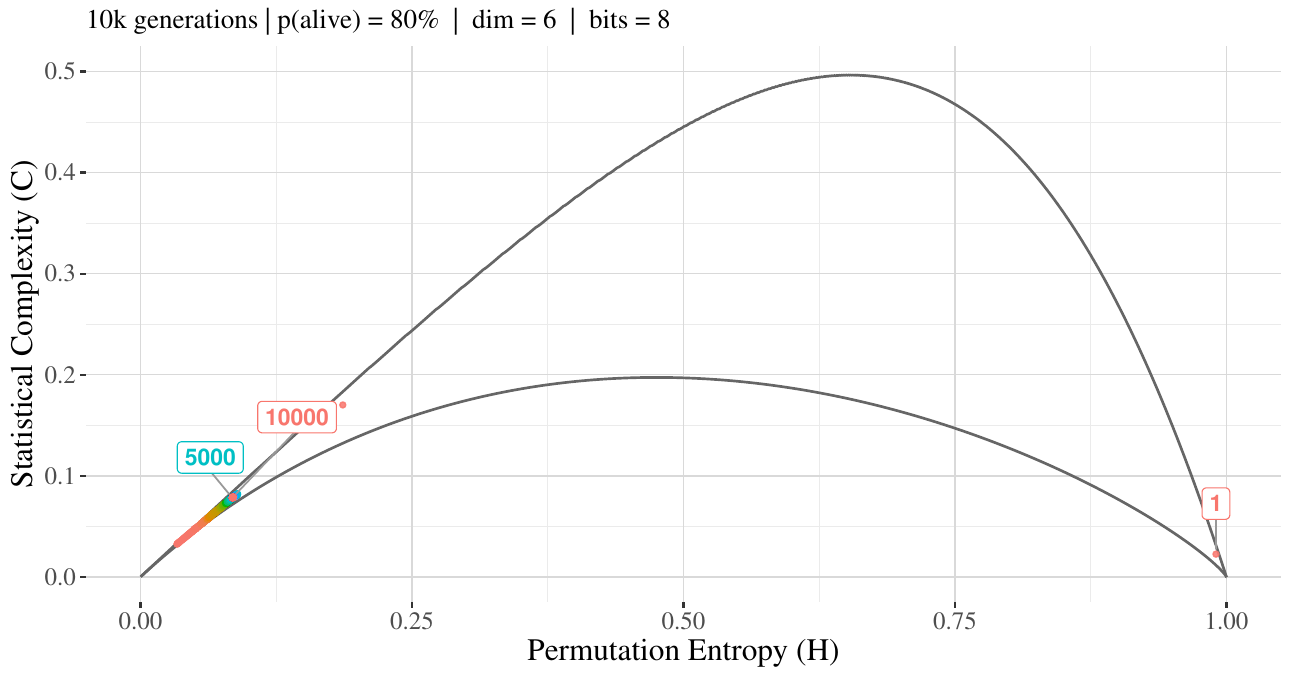}
    \caption{$\rho_0=80\%$}
    \label{fig:initial_80_8b}
\end{subfigure}
\caption{entropy--complexity trajectories for different initial fractions of live cells in Conway's Game of Life. All simulations use $10000$ generations, embedding dimension $D=6$, and a 8-bit Peano-Hilbert encoding.}
\label{fig:hxc-dim6-8bit}
\end{figure*}

\begin{figure*}[]
\centering
\begin{subfigure}[b]{0.48\textwidth}
    \centering
    \includegraphics[width=\textwidth]{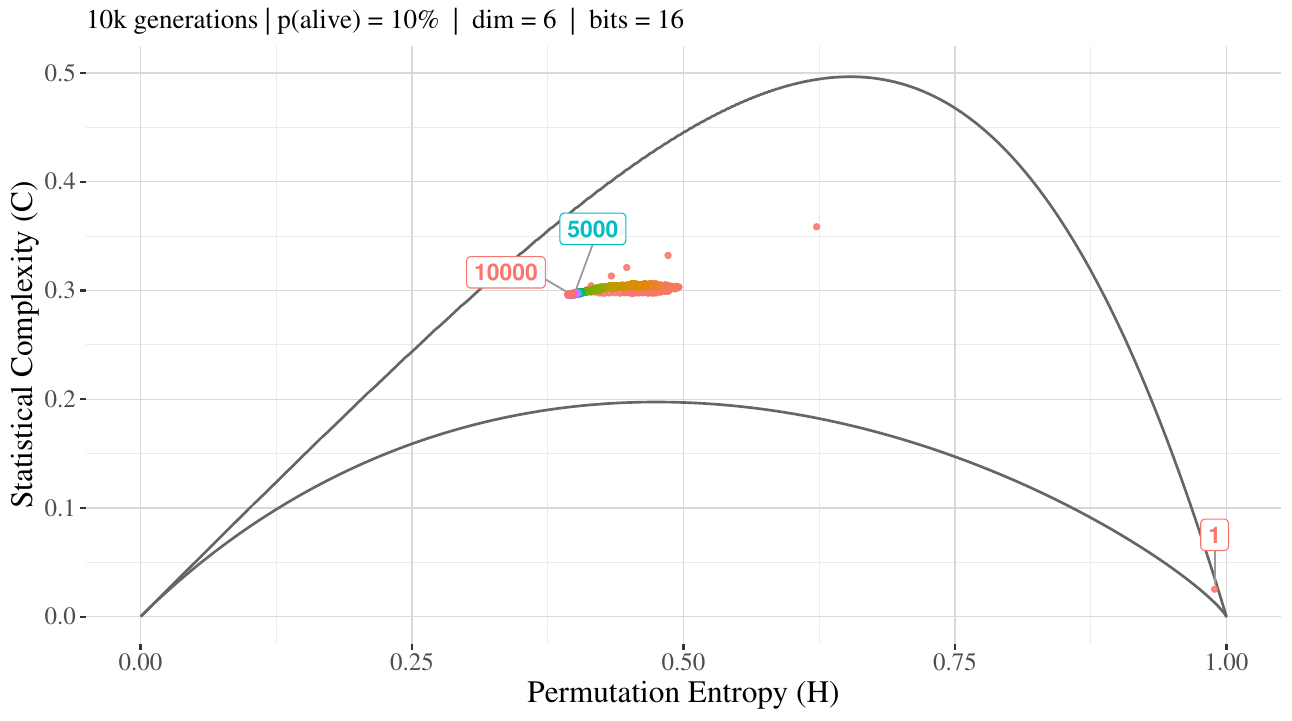}
    \caption{$\rho_0=10\%$}
    \label{fig:initial_10_16b}
\end{subfigure}
\hfill
\begin{subfigure}[b]{0.48\textwidth}
    \centering
    \includegraphics[width=\textwidth]{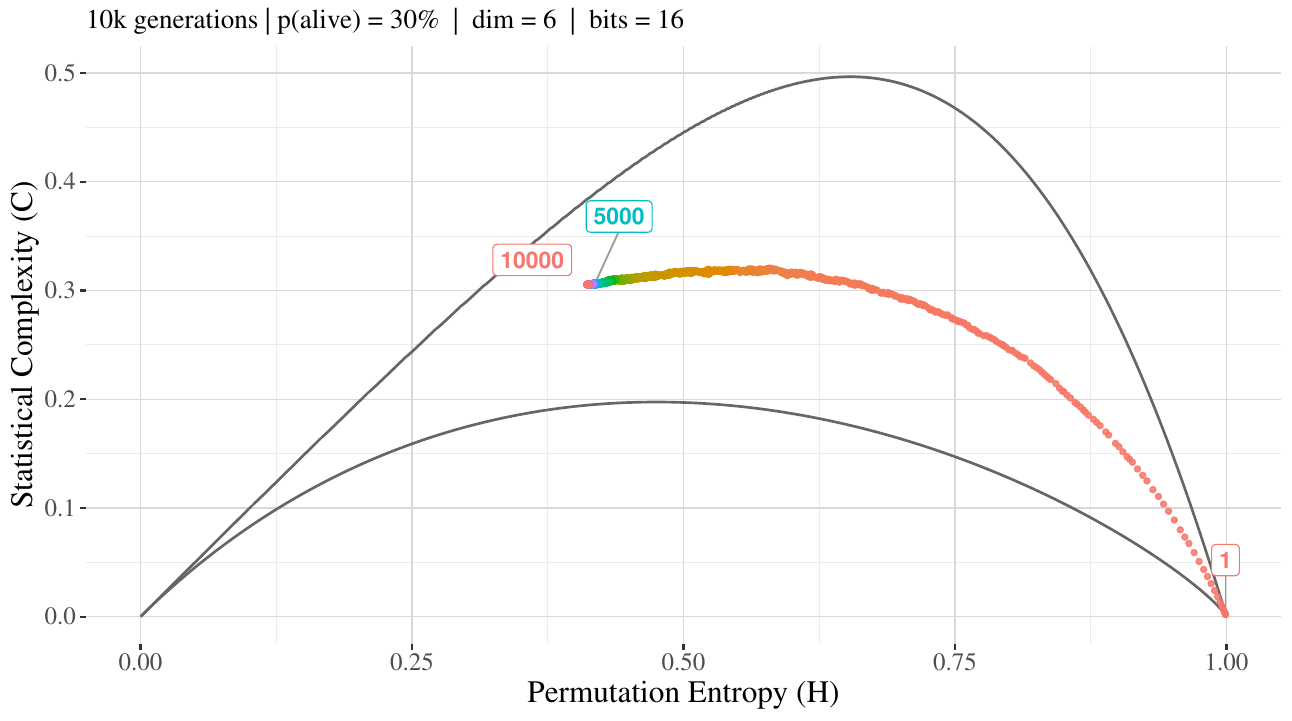}
    \caption{$\rho_0=30\%$}
    \label{fig:initial_30_16b}
\end{subfigure}
\vspace{0.5em}
\begin{subfigure}[b]{0.48\textwidth}
    \centering
    \includegraphics[width=\textwidth]{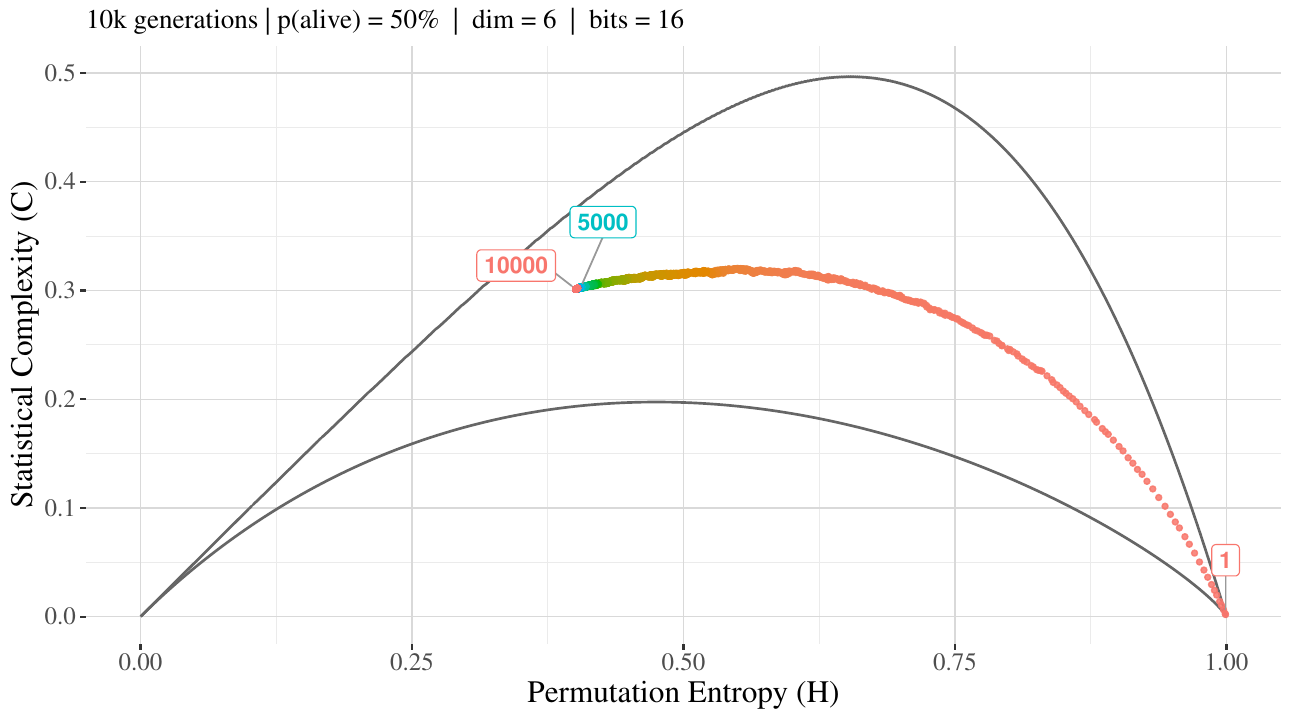}
    \caption{$\rho_0=50\%$}
    \label{fig:initial_50_16b}
\end{subfigure}
\hfill
\begin{subfigure}[b]{0.48\textwidth}
    \centering
    \includegraphics[width=\textwidth]{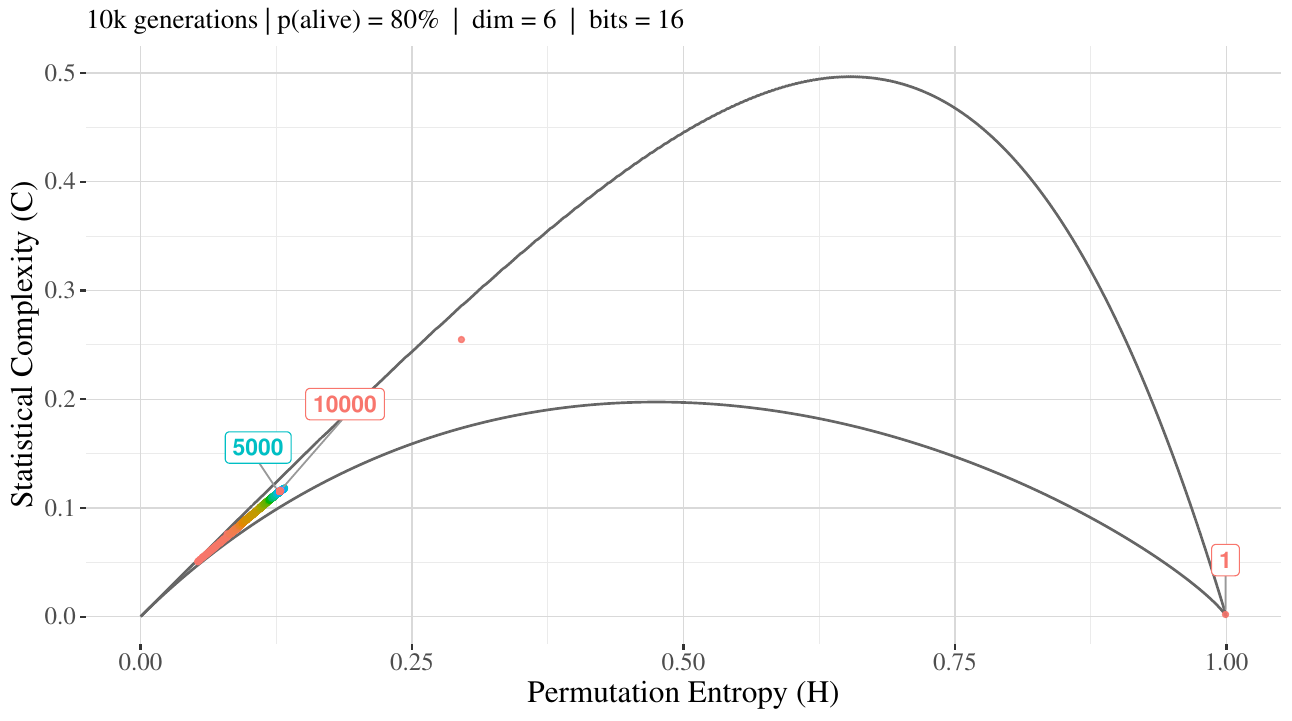}
    \caption{$\rho_0=80\%$}
    \label{fig:initial_80_16b}
\end{subfigure}
\caption{entropy--complexity trajectories for different initial fractions of live cells in Conway's Game of Life. All simulations use $10000$ generations, embedding dimension $D=6$, and a 16-bit Peano-Hilbert encoding.}
\label{fig:hxc-dim6-16bit}
\end{figure*}

\begin{figure*}[]
\centering
\begin{subfigure}[b]{0.48\textwidth}
    \centering
    \includegraphics[width=\textwidth]{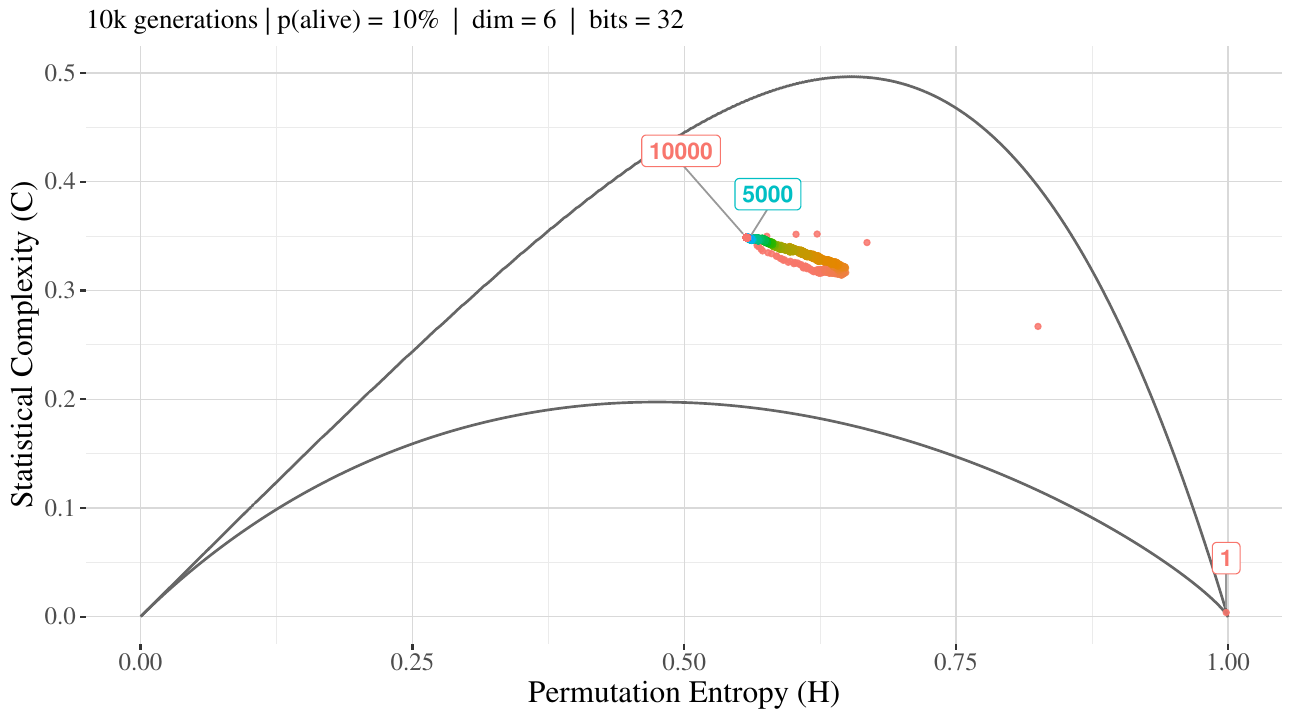}
    \caption{$\rho_0=10\%$}
    \label{fig:initial_10_32b}
\end{subfigure}
\hfill
\begin{subfigure}[b]{0.48\textwidth}
    \centering
    \includegraphics[width=\textwidth]{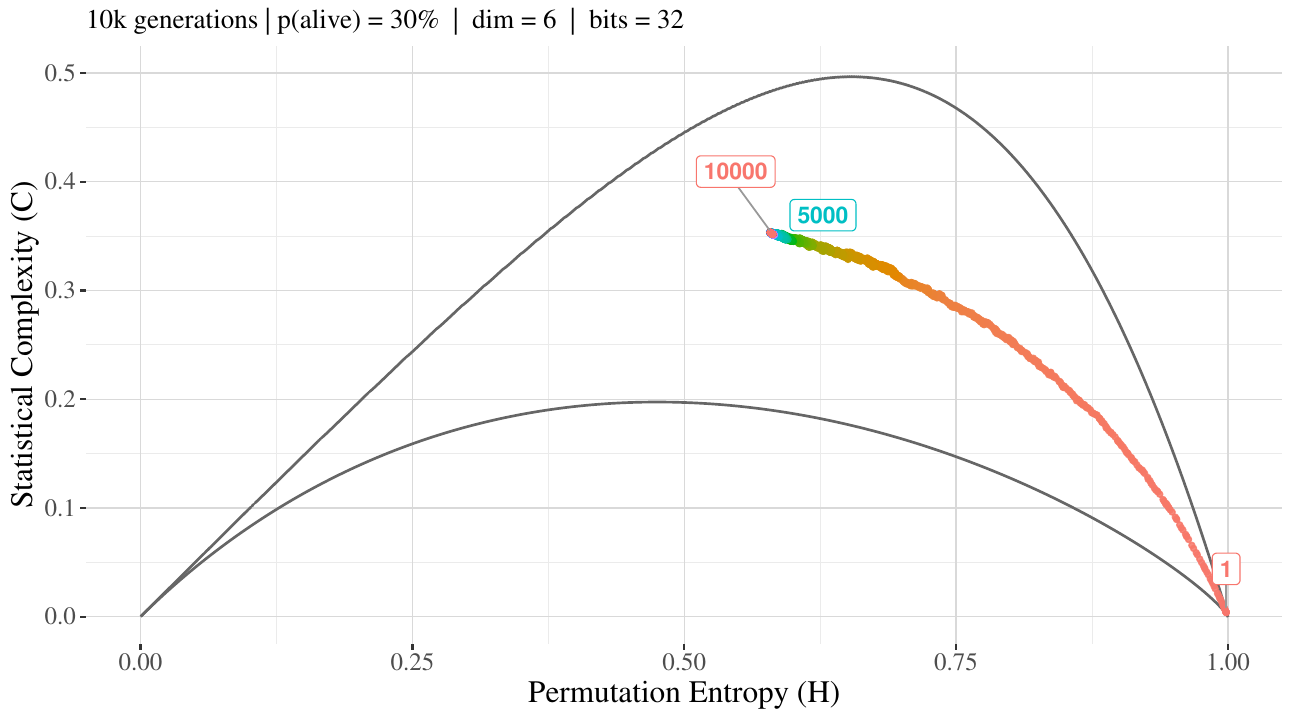}
    \caption{$\rho_0=30\%$}
    \label{fig:initial_30_32b}
\end{subfigure}
\vspace{0.5em}
\begin{subfigure}[b]{0.48\textwidth}
    \centering
    \includegraphics[width=\textwidth]{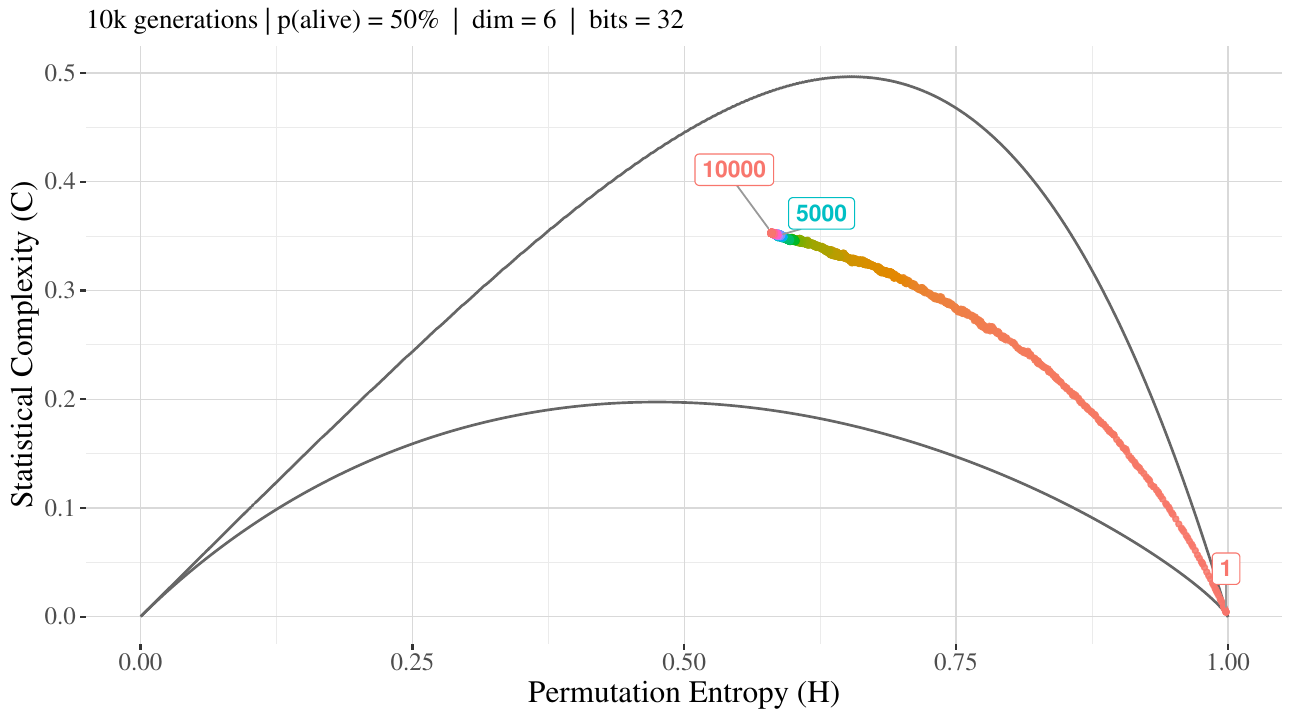}
    \caption{$\rho_0=50\%$}
    \label{fig:initial_50_32b}
\end{subfigure}
\hfill
\begin{subfigure}[b]{0.48\textwidth}
    \centering
    \includegraphics[width=\textwidth]{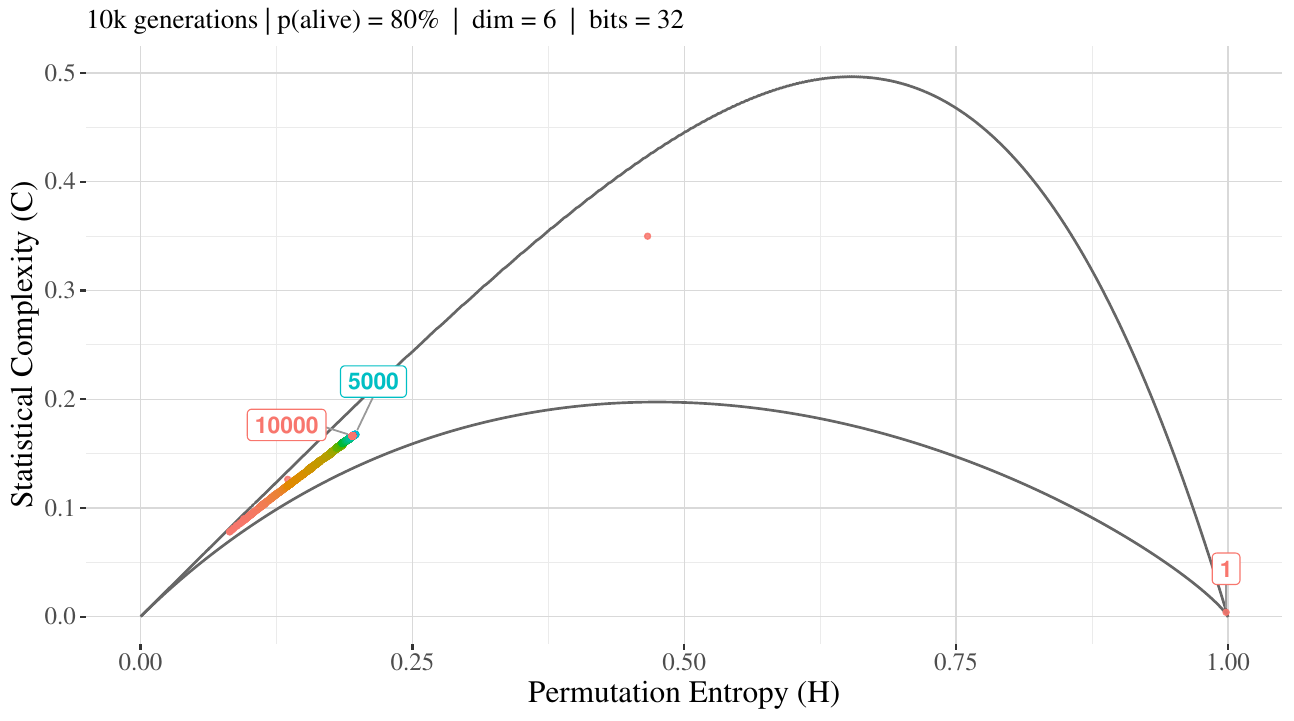}
    \caption{$\rho_0=80\%$}
    \label{fig:initial_80_32b}
\end{subfigure}
\caption{entropy--complexity trajectories for different initial fractions of live cells in Conway's Game of Life. All simulations use $10000$ generations, embedding dimension $D=6$, and a 32-bit Peano-Hilbert encoding.}
\label{fig:hxc-dim6-32bit}
\end{figure*}

We first observe trajectories of the GoL dynamics in the entropy--complexity plane for an embedding dimension of $D=6$, considering three bit lengths ($b=2^3,2^4,2^5$) and four initial densities ($\rho_0\%=10,30,50,80$). 
Figure~\ref{fig:hxc-dim6-8bit} illustrates the trajectories obtained with $b=8$. 
Figure~\ref{fig:hxc-dim6-16bit} presents the corresponding trajectories for $b=16$, while Figure~\ref{fig:hxc-dim6-32bit} shows the results for $b=32$.

When the majority of cells are initially alive 
($\rho_0=80\%$, 
Figures~\ref{fig:initial_80_8b}, \ref{fig:initial_80_16b} and \ref{fig:initial_80_32b}), 
the trajectory remains concentrated in the lower-left region of the $H\times C$ plane. This position indicates a regime characterized by low permutation entropy and low statistical complexity, corresponding to highly regular temporal patterns with limited diversity in their local structures. 
A different behavior is observed when the initial configuration contains only $\rho_0=10\%$ of live cells; cf.\ Figures~\ref{fig:initial_10_8b}, \ref{fig:initial_10_16b}, and \ref{fig:initial_10_32b}. 
Although the trajectories remain concentrated within a relatively well-defined region of the entropy--complexity plane, increasing the bit depth of the Peano-Hilbert encoding leads to higher values of both entropy and statistical complexity. This indicates that a finer encoding captures additional information about the underlying spatiotemporal structure of the GoL dynamics, resulting in a displacement of the trajectories toward regions of higher entropy and complexity.

Notably, for initial configurations with $30\%$ 
(Figures~\ref{fig:initial_30_8b}, \ref{fig:initial_30_16b}, and~\ref{fig:initial_30_32b})
and $50\%$ 
(Figures~\ref{fig:initial_50_8b}, \ref{fig:initial_50_16b}, and~\ref{fig:initial_50_32b})
live cells, the trajectories follow a well-defined curve in the entropy--complexity plane, eventually reaching a region whose location varies with the Peano-Hilbert encoding bit depth. 
This behavior suggests that, for these initial conditions, the dynamics exhibit more structured trajectories, while the encoding resolution influences the location of the resulting steady state point.

\subsection{Convergence to the Steady State}

Table~\ref{tab:results} summarizes the 
steady states for the 
Permutation Entropy ($H^*$) and 
Statistical Complexity ($C^*$), 
convergence step ($t$), 
and coefficient of determination ($R^2$) of the model defined in Eq.~\eqref{eq:log(dt)} obtained for different initial fractions of live cells ($\rho_0$), embedding dimensions, and Peano-Hilbert bit resolutions.

\begin{table}[]
\caption{Permutation Entropy ($H^*$) and Statistical Complexity ($C^*$) steady states, half-live ($t_{1/2} $), and coefficient of determination ($R^2$) for different initial live cell fractions ($\rho_0$), embedding dimensions (D), and Peano-Hilbert bit resolutions.}
\label{tab:results}
\begin{tabular*}{\tblwidth}{@{}*{14}R@{}}
\toprule
\multirow{2}{*}{$\rho_0$} & \multirow{2}{*}{D} 
& \multicolumn{4}{c}{8} 
& \multicolumn{4}{c}{16} 
& \multicolumn{4}{c}{32} \\
\cmidrule(lr){3-6} \cmidrule(lr){7-10} \cmidrule(lr){11-14}
& & $H^*$ & $C^*$ & $t_{1/2} $ & $R^2$
& $H^*$ & $C^*$ & $t_{1/2} $ & $R^2$
& $H^*$ & $C^*$ & $t_{1/2} $ & $R^2$ \\
\midrule

\multirow{2}{*}{10\%}
& 5 
& 0.295 & 0.219 & 1119 & 0.880
& 0.424 & 0.277 & 1113 & 0.868
& 0.593 & 0.299 & 1172 & 0.876 \\

& 6 
& 0.272 & 0.225 & 1088 & 0.883
& 0.396 & 0.297 & 1161 & 0.871
& 0.560 & 0.348 & 1163 & 0.880 \\

\midrule

\multirow{2}{*}{30\%}
& 5 
& 0.303 & 0.225 & 922 & 0.917
& 0.437 & 0.282 & 941 & 0.917
& 0.611 & 0.299 & 1035 & 0.913 \\

& 6 
& 0.280 & 0.230 & 925 & 0.920
& 0.408 & 0.303 & 949 & 0.925
& 0.579 & 0.351 & 1059 & 0.906 \\

\midrule

\multirow{2}{*}{50\%}
& 5 
& 0.304 & 0.225 & 1080 & 0.903
& 0.436 & 0.281 & 1175 & 0.887
& 0.610 & 0.300 & 1184 & 0.872 \\

& 6 
& 0.281 & 0.231 & 1089 & 0.903
& 0.407 & 0.304 & 1179 & 0.885
& 0.577 & 0.351 & 1186 & 0.883 \\

\midrule

\multirow{2}{*}{80\%}
& 5 
& 0.101 & 0.089 & 1199 & 0.807
& 0.152 & 0.128 & 922 & 0.887
& 0.223 & 0.174 & 1140 & 0.855 \\

& 6 
& 0.093 & 0.085 & 1141 & 0.828
& 0.139 & 0.124 & 974 & 0.888
& 0.205 & 0.173 & 1085 & 0.871 \\

\bottomrule
\end{tabular*}
\end{table}

It is observed that $H^*$ and $C^*$ generally increase as the number of bits increases, regardless of the initial alive-cell fraction or embedding dimension. 
For example, for $\rho_0=10\%$ and embedding dimension~$5$, $H^*$ increases from $0.295$ ($8$ bits) to $0.424$ ($16$ bits) and $0.593$ ($32$ bits), while $C^*$ increases from $0.219$ to $0.277$ and $0.299$, respectively.
A similar trend is observed for the other initial conditions. 
This behavior indicates that increasing the bit resolution produces a finer representation of the spatial configuration, capturing more information about the underlying spatiotemporal structure. 
The effect is particularly pronounced for entropy, whereas the increase in statistical complexity is comparatively moderate.

The initial fraction of live cells strongly influences the dynamical regime. 
For $\rho_0 \in \{10,30,50\}$, the resulting values of $H^*$ and $C^*$ remain within relatively similar ranges for a given bit resolution. 
In contrast, $\rho_0=80\%$ consistently produces substantially lower values of both measures. 
For example, at $32$ bits and embedding dimension~$5$, $H^*$ decreases from approximately $0.59\rightarrow0.61$ for $\rho_0 \in \{10,30,50\}$ to $0.223$ for $\rho_0=80\%$, while $C^*$ decreases from approximately $0.30 \rightarrow 0.174$. 
This indicates that the high-density initial configuration leads to a qualitatively different dynamical regime, characterized by lower entropy and statistical complexity.

Increasing the embedding dimension from $5$ to $6$ produces a consistent but relatively small reduction in $H^*$ and increase in $C^*$. 
For example, at $\rho_0=50\%$ and $32$ bits, $H^*$ decreases from $0.610$ to $0.577$, whereas $C^*$ increases from $0.300$ to $0.351$. 
This behavior suggests that increasing the embedding dimension captures additional structural regularities in the dynamics, resulting in a representation characterized by lower entropy and higher statistical complexity.

The convergence results provide further insight into these differences. 
The half-life $t_{1/2}$ varies considerably with the initial live-cell fraction and bit resolution. 
For $\rho_0=30\%$, the system reaches the steady state relatively early, with $t_{1/2}$ ranging from $922$ to $1059$ across the considered configurations. 
For $\rho_0=10\%$, the convergence is usually achieved later, ranging from $1088$ to $1172$. Similarly, the $50\%$ configurations exhibit half-lives between $1080$ and $1186$. 
The $80\%$ condition does not uniformly converge faster than the other configurations: although some cases exhibit relatively small values of $t_{1/2}$, the half-life ranges from $922$ to $1199$ across the different resolutions and embedding dimensions. 
Therefore, the updated results do not support a monotonic relationship between initial live-cell fraction and convergence speed.

These results contrast with the generation lengths adopted in previous studies of complexity and entropy in GoL (Table~\ref{tab:generation-lit}).
The surveyed works use $L$ between $7$ and $400$, well below the half-lives observed here ($t_{1/2} \in [922, 1199]$), so their simulations likely capture transient dynamics rather than the steady state.

\begin{table}[]
\caption{Generation length used in the literature.}\label{tab:generation-lit}
\begin{tabular*}{\tblwidth}{@{}LC@{}}
\toprule
Work & Generation length L \\ 
\midrule
\citealp{Speck1978} & 400 \\
\citealp{pan2010emergencesymmetrynewtype} & 7 \\
\citealp{ContrerasReyes2021Lerch} & 200 \\
\citealp{Kharazmi2023deng} & 200 \\
\citealp{ConwaysGameo2024l} & 100 \\
\bottomrule
\end{tabular*}
\end{table}

Finally, the $R^2$ values indicate that the logarithmic-distance fits used to characterize convergence provide a strong approximation across all configurations, with values ranging from $0.807$ to $0.920$. 
The highest values are generally observed for the $\rho_0=30\%$ initial condition, reaching $R^2=0.925$ for $b=16$ and embedding dimension~$6$. 
In contrast, the lowest values occur primarily for the $\rho_0=80\%$ initial condition, particularly at $b=8$, where $R^2=0.807$ and $0.828$ are obtained for embedding dimensions~$5$ and $6$, respectively. 
These results indicate that the exponential-decay model consistently explains a substantial fraction of the variation in $\log(d_t)$ across the considered cases.

\begin{figure*}[]
\centering
\begin{subfigure}[b]{0.48\textwidth}
    \centering
    \includegraphics[width=\textwidth]{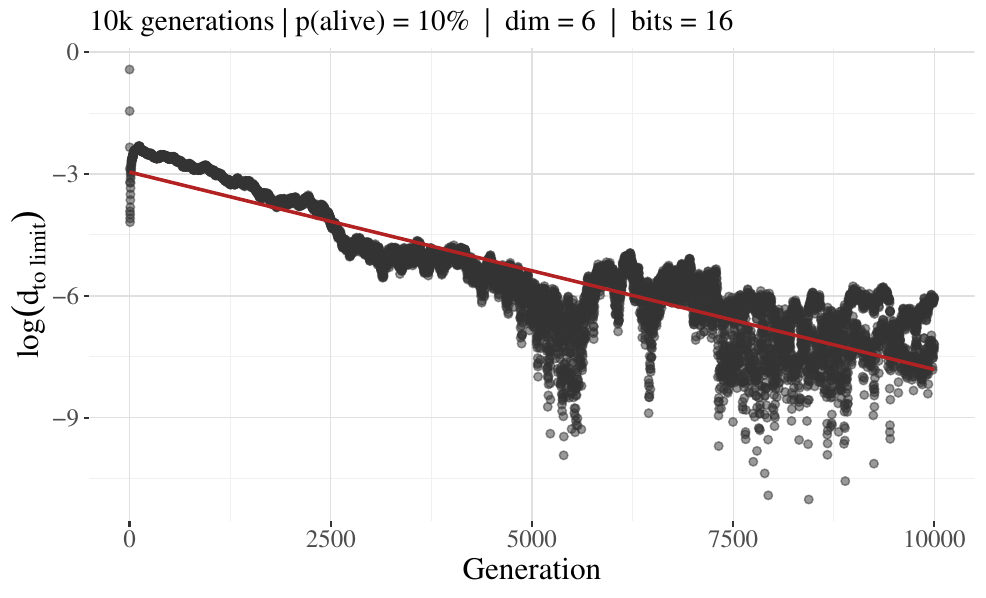}
    \caption{$\rho_0=10\%$}
    \label{fig:conv_10}
\end{subfigure}
\hfill
\begin{subfigure}[b]{0.48\textwidth}
    \centering
    \includegraphics[width=\textwidth]{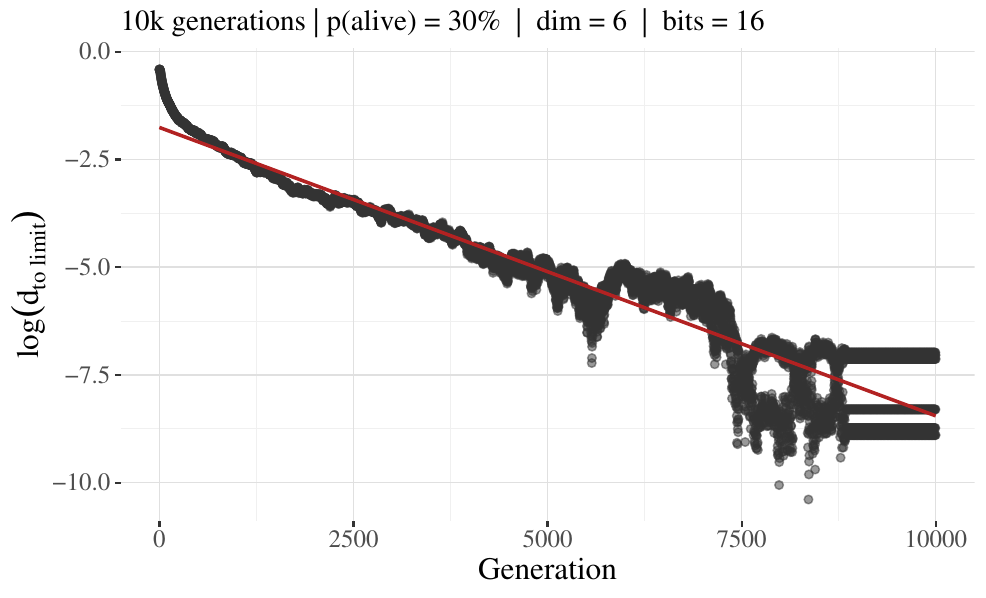}
    \caption{$\rho_0=30\%$}
    \label{fig:conv_30}
\end{subfigure}
\vspace{0.5em}
\begin{subfigure}[b]{0.48\textwidth}
    \centering
    \includegraphics[width=\textwidth]{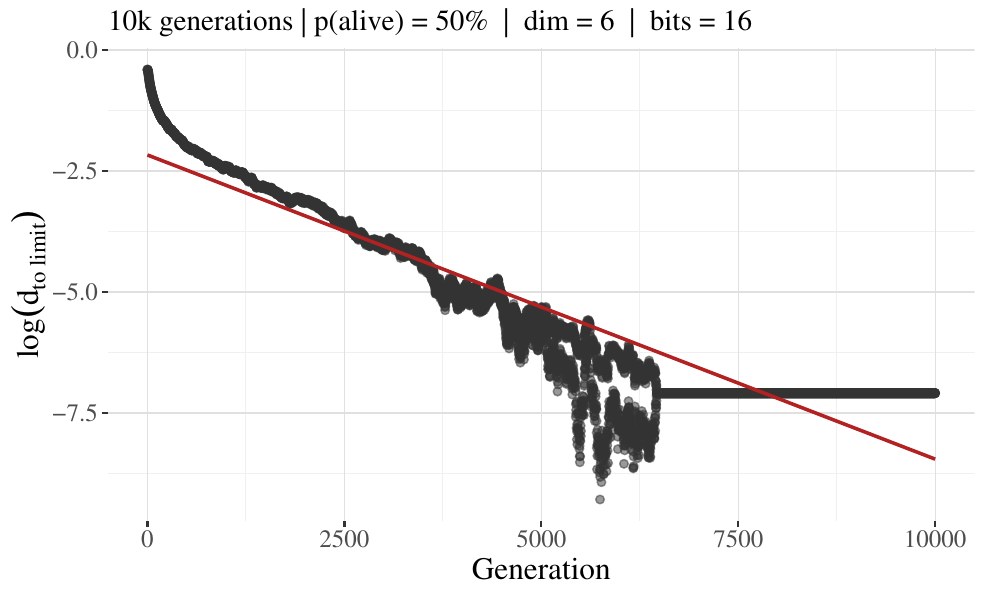}
    \caption{$\rho_0=50\%$}
    \label{fig:conv_50}
\end{subfigure}
\hfill
\begin{subfigure}[b]{0.48\textwidth}
    \centering
    \includegraphics[width=\textwidth]{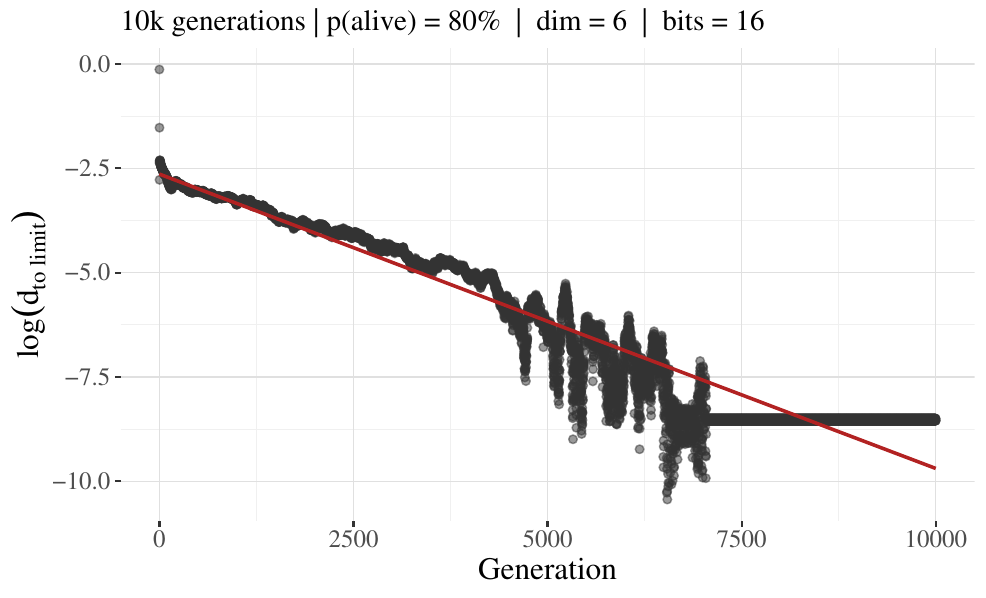}
    \caption{$\rho_0=80\%$}
    \label{fig:conv_80}
\end{subfigure}
\caption{Convergence to the steady state $(H^*, C^*)$ across initial density levels, for embedding dimension $D=dim=6$ and bit-depth $b=16$. Each panel shows $\log d_t$ (grey points) versus generation, with the fitted log-linear regression (red line) used to estimate the convergence rate $r$. (a) $\rho_0=10\%$; (b) $\rho_0=30\%$; (c) $\rho_0=50\%$; (d) $\rho_0=80\%$. Higher initial densities reach the discretization-limited plateau earlier, while lower densities exhibit greater variance in the saturated regime.}
\label{fig:convergence-density}
\end{figure*}

Figure~\ref{fig:convergence-density} shows the convergence trajectories (log-distance to the steady state, $\log d_t$) for embedding dimension~$6$ and bit-depth $b=16$, across four initial density levels: $\rho_0 \in \{10\%, 30\%, 50\%, 80\%\}$. 
In the four cases, the system exhibits an initial phase of approximately log-linear decay, consistent with exponential convergence toward the reference point $(H^*, C^*)$, followed by saturation into a noisy plateau. 
This plateau reflects the discretization limit imposed by the $b=16$ resolution, producing the characteristic banded structure visible at later generations.

The onset of saturation varies with initial density: higher densities ($\rho_0=50\%$, $p=80\%$) reach the plateau earlier (around generation $6000$--$7000$) than lower densities ($\rho_0=10\%$, $\rho_0=30\%$), which continue decaying past generation $7500$. 
The $\rho_0=10\%$ condition additionally shows markedly higher variance in the saturated regime, with intermittent excursions to substantially lower $d_t$ values, likely reflecting stochastic fluctuations associated with sparse initial populations (near-extinction events or small-scale pattern collapse).

Because the log-linear regression is fit across the full trajectory, the estimated convergence rate $r$ (and derived half-life $t_{1/2} = \ln(2)/r$) may be biased toward faster apparent convergence for configurations that saturate earlier, since post-saturation points contribute residual downward pull to the fit. 
We note this as a methodological consideration for rate estimation in Section~\ref{sec:convergence}.

\section{Final Considerations}

In this work, we presented a methodology for characterizing the structural information of Conway's Game of Life (GoL) through the entropy--complexity ($H \times C$) plane. 
The proposed approach converts each generation of the cellular automaton into a time series by traversing the lattice via a Peano-Hilbert space-filling curve and mapping the resulting bit sequences to decimal values. 
Each generation's time series is then transformed into ordinal patterns, 
from which the Shannon permutation entropy $H$ and Jensen-Shannon statistical complexity $C$ are estimated, allowing the CA dynamic to be tracked as a trajectory in the $H \times C$ plane.

Compared to previous work, the cellular automaton configurations explored here are substantially larger, both in lattice size ($1024 \times 1024$) and in the number of generations simulated (up to $10000$), providing a more detailed view of long-term dynamical behavior. 
As expected of a chaotic system, the GoL exhibits strong sensitivity to initial conditions: varying the initial density of live cells $\rho_0$ produces markedly different trajectories in the $H \times C$ plane, both in terms of the path taken and the point of convergence.

Across all tested configurations, trajectories consistently converge toward a stable region of the plane, indicating that despite the CA's sensitivity to initial conditions, its long-run statistical complexity settles into a well-defined steady state $(H^*, C^*)$. 
This convergence follows an approximately exponential decay in the early stages of the simulation, after which it saturates into a plateau imposed by the finite bit-depth resolution used in the discretization step. 
The onset of this saturation, and the apparent rate of convergence, vary with the initial density $\rho_0$: configurations with higher initial densities tend to reach the plateau earlier, while lower-density configurations show greater variance in the saturated regime, likely due to stochastic effects associated with sparser initial populations.

These results suggest that the entropy--complexity plane, 
combined with the proposed Peano-Hilbert encoding and conversion from binary to integer (bit-depth encoding), 
provides a robust and scalable framework for characterizing the dynamical regime of cellular automata, capturing both the transient and steady state behavior of the system. 
As future work, we intend to extend this methodology to the two-dimensional Ising model, investigating whether the same Peano-Hilbert encoding and $H \times C$ plane framework can characterize its order-disorder phase transition and identify signatures of criticality analogous to the convergence regimes observed here for the Game of Life.

\appendix
\section{Thinning the parametric space}\label{app:ParametricSpace}

To identify the parameter configurations most representative of the experiments, the proposed methodology was evaluated on a smaller lattice of $2^8 \times 2^8$ ($256\times256$) cells using a broader range of parameter values, cf.\ Table~\ref{tab:TotalParameters}. 
This preliminary analysis was designed to identify parameter configurations that provide sufficiently informative representations of the cellular automaton dynamics, while reducing the number of configurations for the subsequent experiments.

\begin{table}[]
\caption{Parameter values explored in the preliminary experiments.}
\label{tab:TotalParameters}
\begin{tabular*}{\tblwidth}{@{}LLL@{}}
\toprule
Category & Parameter & Values \\
\midrule
\multirow{2}{*}{CA Simulation} & Number of generations & $\{3000, 5000, 10000\}$ \\
 & Initial living-cells ($\rho_0$) & $\{10, 30, 50, 80\}$ \\
Ordinal Patterns & Embedding dimension & $\{3, 4, 5, 6\}$ \\
Peano-Hilbert & Bits & $\{2, 4, 8, 16, 32\}$ \\
\bottomrule
\end{tabular*}
\end{table}

\begin{table}[]
\caption{Time series size for a $256\times256$ cellular automaton lattice as a function of bit-depth $b$, computed as $N = 256^2 / b$, alongside the number of possible encoded states, $2^b$.} \label{tab:ca_timeseries_size}
\begin{tabular*}{\tblwidth}{@{}LLL@{}}
\toprule
$b$ (bits) & Time Series Size & \# states ($2^b$) \\ %
\midrule
2  & 32768 & $4.00\times10^{0}$ \\
4  & 16384 & $1.60\times10^{1}$ \\
8  & 8192  & $2.56\times10^{2}$ \\
16 & 4096  & $6.55\times10^{4}$ \\
32 & 2048  & $4.29\times10^{9}$ \\
\bottomrule
\end{tabular*}
\end{table}

A two-stage factorial sweep was conducted. 
First, 12 independent CA simulations (randomly initialized) were generated from a full factorial combination of 
the number of generations ($T\in\{3000, 5000, 10000\}$) 
and the initial density of live cells $\rho_0\in\{10\%, 30\%, 50\%, 80\%\}$. 
Each simulation was then analyzed across a second factorial sweep over the embedding dimension  $D\in\{3, 4, 5, 6\}$ 
and the bit-depth encoding $b\in\{2, 4, 8, 16, 32\}$, 
yielding $4\times5 = 20$ analysis configurations per simulation. 
In total, this resulted in $12\times20 = 240$ experiments. 

For each configuration, the entropy $H$ and statistical complexity $C$ were computed throughout the generations, forming a trajectory in the entropy--complexity plane, as described in section \ref{sec:proposedmethod}. 
To characterize the long-term behavior of each trajectory, the mean of its final portion, defined by the parameter $\upsilon$, was computed to identify the steady state point $(H^*,C^*)$ of the plane toward which the system converges (section~\ref{sec:convergence}). 
This tail-averaging procedure provides an estimate of the steady state behavior while reducing the influence of transient fluctuations. 
An appropriate value of $\upsilon$ was first determined for each simulation duration, and the selected values were subsequently used to evaluate the overall effects of the remaining parameters.

\subsection{Estimation of $\upsilon$}\label{app:TauEstimation}

The choice of $\upsilon$ may affect the estimated steady state point and, consequently, the estimated decay rate, therefore a sensitivity analysis was performed to assess the robustness of this parameter. 
Among the evaluated initial densities, the configuration with $\rho_0=50\%$ and $\rho_0=30\%$ exhibited the longest trajectory before reaching the convergence regime (very distinct from $10\%$ and $80\%$), as seen in~Figure \ref{fig:hxc-dim6-8bit}. 
Thus, both initial configurations were selected for the subsequent sensitivity analysis of $\upsilon$.

The $\upsilon$ value was initially selected from the stationary portion of the entropy--complexity trajectory. 
To assess the sensitivity of the resulting estimates to this choice, $\upsilon$ was subsequently varied as a fraction of the total simulation duration. 
The longest simulation, corresponding to $T=10000$ generations, was considered, and candidate values of $\upsilon$ ranging from $10\%$ to $80\%$ of the trajectory length were evaluated, corresponding to $\upsilon\in\{1000, 2000, 4000, 6000, 8000\}$ generations. 
For each candidate window, the decay rate $r$ and the corresponding half-life $t_{1/2}= \ln(2) / r$ were re-estimated and compared with the mean half-life obtained across all evaluated windows.

\begin{figure}[]
\centering
\includegraphics[width=0.9\columnwidth]{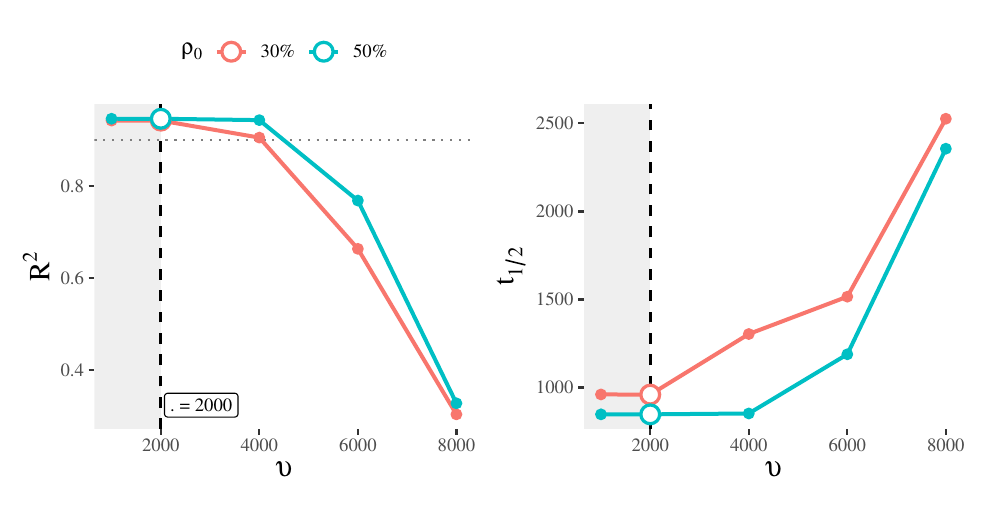}
\caption{Sensitivity of the convergence-rate estimate to the tail window size $\upsilon$, for initial living-cell densities $\rho_0=30\%$ and $\rho_0=50\%$ (embedding dimension $D=6$, bit depth $=8$). \textbf{Left:} fit quality ($R^2$) of the log-linear decay model as a function of $\upsilon$; the dotted horizontal line marks $R^2=0.9$. \textbf{Right:} corresponding half-life estimate $t_{1/2}$. In both panels, the shaded region indicates the stable range $\upsilon \leq 2000$, in which the estimated limit point $(H^*, C^*)$ is insensitive to the tail window size for both densities; the dashed vertical line and open circles mark the adopted value $\upsilon=2000$, used for all subsequent experiments. Beyond this range, the estimate degrades progressively as the tail window begins to incorporate transient, pre-convergence dynamics.}
\label{fig:ntail_sensitivity_lc}
\end{figure}

Figure~\ref{fig:ntail_sensitivity_lc} presents the sensitivity of the convergence-rate estimate to the tail window size $\upsilon$ for two initial living-cell densities ($\rho_0=30\%$ and $\rho_0=50\%$), at fixed embedding dimension $D=6$ and bit depth $=8$. 
The left panel shows the fit quality ($R^2$) of the log-linear decay model as a function of $\upsilon$, while the right panel shows the corresponding half-life estimate $t_{1/2}$. 
For $\upsilon \leq 2000$, both configurations yield stable and nearly identical estimates ($R^2 \geq 0.94$ in both cases), indicating that the limit point $(H^*, C^*)$ is reliably estimated within this range regardless of $LC$. 
Beyond $\upsilon=2000$, the two configurations begin to diverge: $\rho_0=30\%$ degrades earlier, with $R^2$ dropping to $0.906$ already at $\upsilon=4000$, 
whereas $\rho_0=50\%$ remains stable up to $\upsilon=4000$ ($R^2=0.944$) before deteriorating more sharply at $\upsilon=6000$. 
This indicates that the size of the stable tail window depends on the specific configuration, and that a single value of $\upsilon$ must be chosen conservatively to remain valid across all configurations evaluated in this study. 
Based on this analysis, we adopt $\upsilon=2000$ (corresponding to $20\%$ final $H\times C$ points) for all subsequent experiments, as it lies within the stable regime for every configuration tested and maximizes the averaging window without incorporating transient, pre-convergence dynamics.

\subsection{Parametric study}

To evaluate the impact of the number of generations on the $H\times C$ trajectories, 
we analyzed the number of generations required to reach the minimum entropy $H$ across $30$ independent simulations, for initial densities $\rho_0 \in \{30\%,\,50\%\}$. 
For each $\rho_0$, each simulation $k = 1,\dots,30$, and each time horizon $T \in \{3000, 5000, 10000\}$, we computed the minimum entropy over the first $T$ generations, $H^{(k)}_{1:T}$, and $t^{*(k)}_{T}$, the generation at which it is first attained:
\begin{equation}
H^{(k)}_{1:T} = \min_{1 \le t \le T} H^{(k)}(t),
\quad
t^{*(k)}_{T} = \operatorname*{arg\,min}_{1 \le t \le T} H^{(k)}(t),
\text{ for } k = 1,\dots,30, \text{ and } T \in \{3000,\,5000,\,10000\}.
\end{equation}

We also determined the steady state $(H^*, C^*)$ for each parameter configuration in Table~\ref{tab:TotalParameters}.

\begin{figure}[]
\centering
\includegraphics[width=0.7\columnwidth]{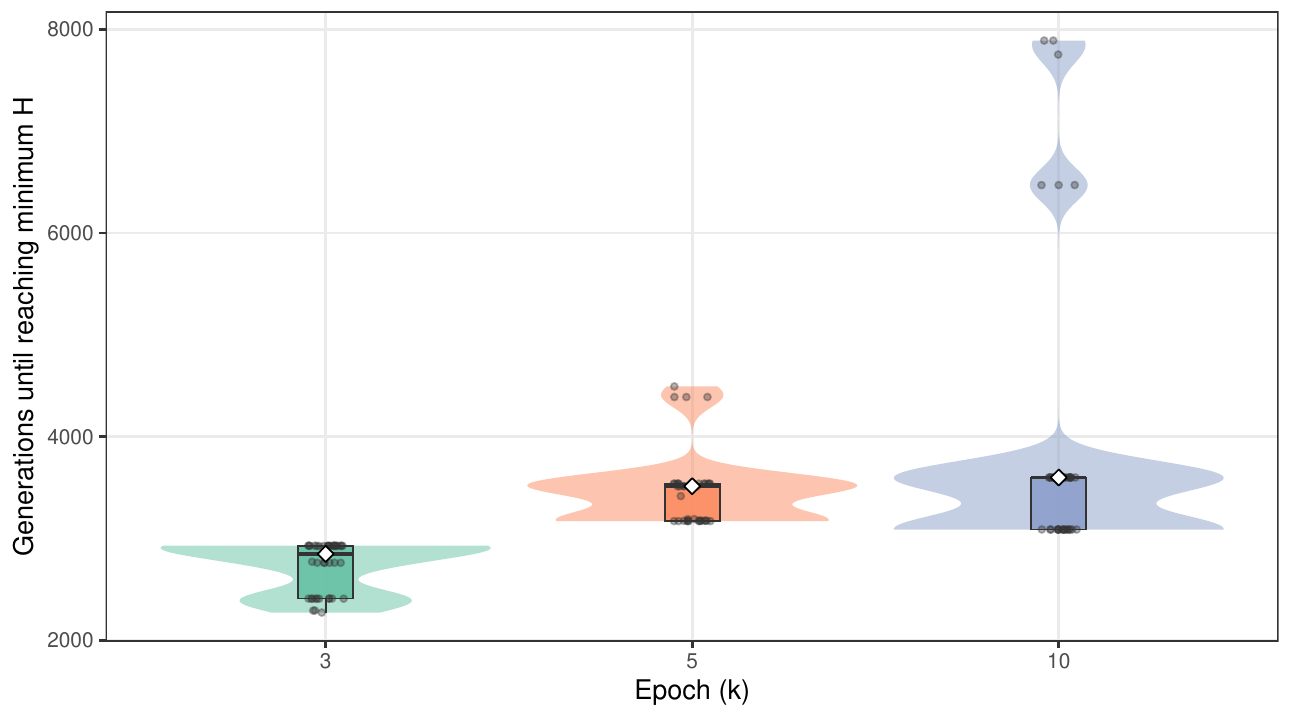}
\caption{Distribution of the number of generations required to reach the minimum entropy ($H$) value, grouped by total simulation length ($3000$, $5000$, and $10000$ generations/epoch), for $\rho_0 \in \{30, 50\}$. Violin plots show the full distribution shape; overlaid boxplots indicate the median and interquartile range; individual jittered points show raw data; white diamonds mark the median. The distribution shifts toward larger values and widens as the total number of generations increases, and both the $5000$ and $10000$ configurations exhibit a secondary cluster of configurations that reach their minimum substantially later than the bulk of the distribution.}
\label{fig:distri}
\end{figure}

The first notable behavior from the results concerns the effect of the total number of simulation generations ($T\in\{3000,5000,10000\}$) on the resulting $H\times C$ trajectories. Longer simulations do not introduce a qualitatively distinct region of the $H\times C$ plane beyond what is already observed in shorter runs, as shown by the comparison of tail-averaged entropy and complexity values. However, Figure~\ref{fig:distri} shows that the number of generations required to reach the minimum entropy value increases and becomes more variable as the total simulation length grows: the median generation at which the minimum is reached rises from approximately $2700$ at $T=3000$ generations to $3400$ at $T=5000$ and $3600$ at $T=10000$, for $\rho_0 \in \{30, 50\}$. 
Moreover, both the $T=5000$ and $T=10000$ configurations exhibit a secondary cluster of configurations that reach their minimum substantially later than the bulk of the distribution (around $4300$-$4400$ generations for $T=5000$, and, more markedly, between $6400$ and $8000$ generations for $T=10000$) indicating that a subset of parameter configurations exhibits qualitatively slower convergence to the minimum entropy as the simulation is allowed to run for longer. 
This later and more variable onset of the entropy minimum implies that shorter runs risk truncating the trajectory before such slow-converging configurations reach their minimum. 
We therefore adopt the longest generation count ($T=10000$) in the subsequent analysis, as it guarantees that any emergent complexity, if present, is more likely to be fully captured than with shorter generation counts.

\begin{figure}[]
\centering
\includegraphics[width=0.7\columnwidth]{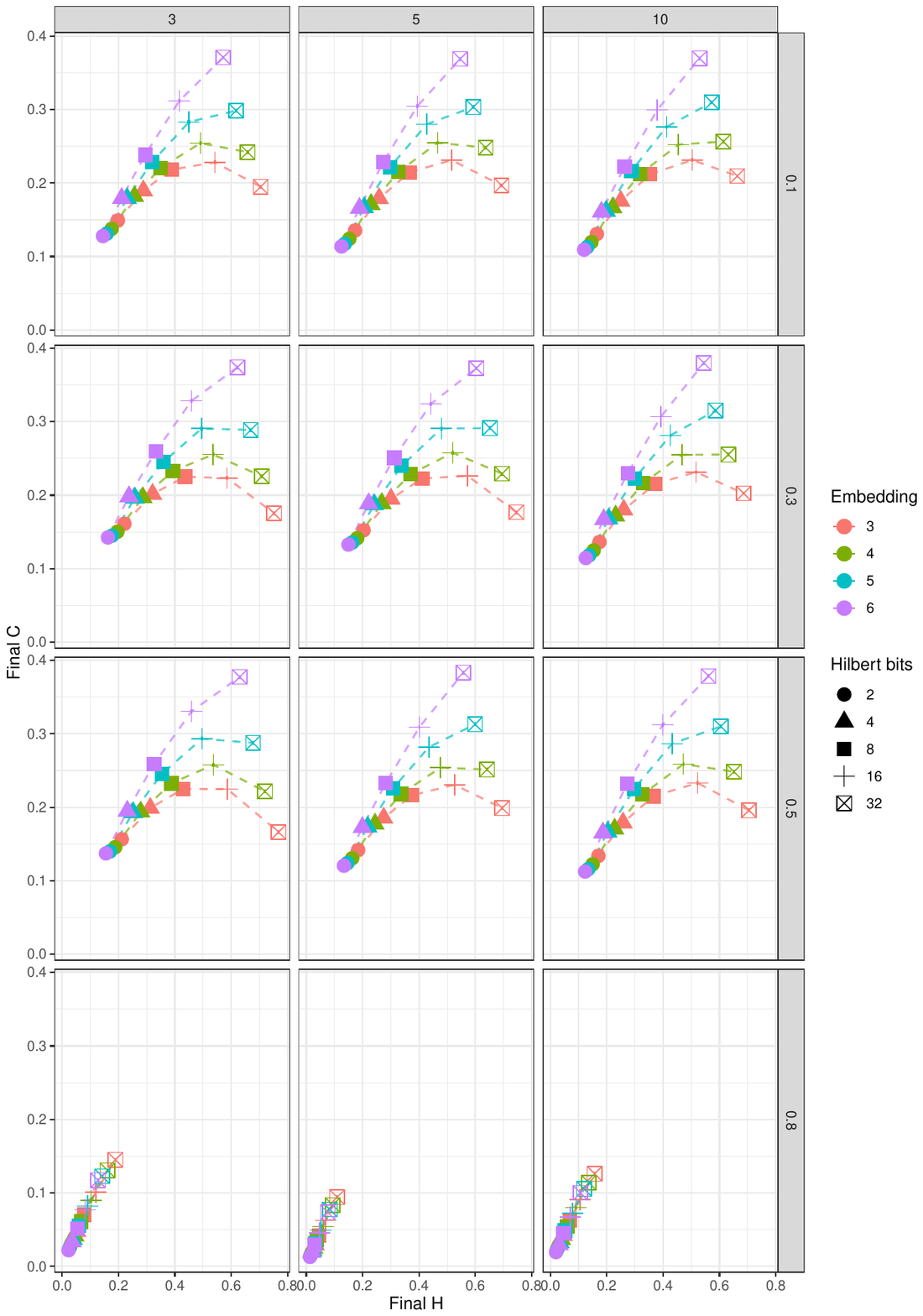}
\caption{Final position in the entropy--complexity ($H\times C$) plane for Conway's Game of Life simulations, as a function of Peano-Hilbert curve bit depth (point shape), embedding dimension (color), initial fraction of live cells ($\rho_0 \in \{10,30,50,80\}$, and number of simulation generations (columns: $3000$, $5000$, and $10000$).}
\label{fig:totalfinalhxc}
\end{figure}

The initial fraction of live cells has a substantial influence on the resulting entropy--complexity trajectories. This effect is expected for the Game of Life, whose subsequent state dynamics is strongly dependent on the initial configuration. Figure~\ref{fig:totalfinalhxc} illustrates that changing $\rho_0$ modifies the region occupied by the final states in the $H\times C$ plane, with particularly distinct behavior observed for the higher-density initial configurations. Rather than selecting a single initial density, we therefore retained the four values $\rho_0 \in \{10,30,50,80\}$ to preserve the range of dynamical regimes observed during the preliminary analysis.

The embedding dimension has a pronounced effect on the estimated statistical complexity. For embedding dimensions $D=3$ and $D=4$, increasing the bit depth tends to produce lower complexity values and a more restricted distribution in the $H\times C$ plane. In contrast, embeddings $D=5$ and $D=6$ preserve a broader range of complexity values and provide a more expressive representation of the dynamical states. This behavior suggests that higher embedding dimensions are better suited to capturing the temporal structure present in the encoded cellular-automaton sequences. Consequently, embeddings $D=5$ and $D=6$ were retained for the subsequent experiments.

The bit depth also produces a clear effect on the distribution of the resulting states. Although smaller values of $b$ produce longer time series, they provide a substantially smaller state universe. For $b=2$, only four distinct encoded states are possible, which limits the amount of information that can be represented by each symbol. This restriction is reflected in Figure~\ref{fig:totalfinalhxc}, where the corresponding configurations are concentrated toward the lower-left region of the entropy--complexity plane. Increasing the bit depth enlarges the state universe and allows the encoding to represent a broader range of local configurations. The results indicate that $b=8$, $b=16$, and $b=32$ provide more informative representations and a broader coverage of the entropy--complexity plane. These values were therefore retained for the subsequent analysis, while $b=2$ and $b=4$ were excluded.

Based on these observations, the preliminary sweep was used to reduce the parameter space for the main experiments. The embedding dimension was restricted to $D\in\{5,6\}$, while the bit depth was restricted to $b\in\{8,16,32\}$. The four initial live-cell fractions and the three simulation durations were retained to account for the sensitivity of the Game of Life dynamics to the initial configuration and simulation horizon. This reduction substantially decreases the number of configurations while preserving the parameter regimes that exhibited the largest variation and most informative coverage of the entropy--complexity plane.

\printcredits

\bibliographystyle{cas-model2-names}

\bibliography{references}

\begin{thebibliography}{34}
\expandafter\ifx\csname natexlab\endcsname\relax\def\natexlab#1{#1}\fi
\providecommand{\url}[1]{\texttt{#1}}
\providecommand{\href}[2]{#2}
\providecommand{\path}[1]{#1}
\providecommand{\DOIprefix}{doi:}
\providecommand{\ArXivprefix}{arXiv:}
\providecommand{\URLprefix}{URL: }
\providecommand{\Pubmedprefix}{pmid:}
\providecommand{\doi}[1]{\href{http://dx.doi.org/#1}{\path{#1}}}
\providecommand{\Pubmed}[1]{\href{pmid:#1}{\path{#1}}}
\providecommand{\bibinfo}[2]{#2}
\ifx\xfnm\relax \def\xfnm[#1]{\unskip,\space#1}\fi
\bibitem[{Bader(2013)}]{Bader2013}
\bibinfo{author}{Bader, M.}, \bibinfo{year}{2013}.
\newblock \bibinfo{title}{Space-Filling Curves: An Introduction with
  Applications in Scientific Computing}.
\newblock \bibinfo{publisher}{Springer}.
\newblock \DOIprefix\doi{10.1007/978-3-642-31046-1}.
\bibitem[{Bandt and Pompe(2002)}]{bandt2002permutation}
\bibinfo{author}{Bandt, C.}, \bibinfo{author}{Pompe, B.}, \bibinfo{year}{2002}.
\newblock \bibinfo{title}{Permutation entropy: A natural complexity measure for
  time series}.
\newblock \bibinfo{journal}{Phys. Rev. Lett.} \bibinfo{volume}{88},
  \bibinfo{pages}{174102}.
\newblock \URLprefix
  \url{https://link.aps.org/doi/10.1103/PhysRevLett.88.174102},
  \DOIprefix\doi{10.1103/PhysRevLett.88.174102}.
\bibitem[{Chagas et~al.(2022)Chagas, Queiroz-Oliveira, Rosso, Ramos, Freitas
  and Frery}]{chagas2022white}
\bibinfo{author}{Chagas, E.T.C.}, \bibinfo{author}{Queiroz-Oliveira, M.},
  \bibinfo{author}{Rosso, O.A.}, \bibinfo{author}{Ramos, H.S.},
  \bibinfo{author}{Freitas, C.G.S.}, \bibinfo{author}{Frery, A.C.},
  \bibinfo{year}{2022}.
\newblock \bibinfo{title}{White noise test from ordinal patterns in the
  {E}ntropy-{C}omplexity plane}.
\newblock \bibinfo{journal}{International Statistical Review}
  \bibinfo{volume}{90}, \bibinfo{pages}{374--396}.
\newblock \URLprefix
  \url{https://onlinelibrary.wiley.com/doi/abs/10.1111/insr.12487},
  \DOIprefix\doi{10.1111/insr.12487}.
\bibitem[{Chopard and Droz(1998)}]{Chopardbook}
\bibinfo{author}{Chopard, B.}, \bibinfo{author}{Droz, M.},
  \bibinfo{year}{1998}.
\newblock \bibinfo{title}{Cellular Automata Modeling of Physical Systems}.
\newblock \bibinfo{publisher}{Springer, New York, NY}.
\newblock \DOIprefix\doi{10.1007/978-0-387-30440-3_57}.
\bibitem[{Contreras-Reyes(2021)}]{ContrerasReyes2021Lerch}
\bibinfo{author}{Contreras-Reyes, J.E.}, \bibinfo{year}{2021}.
\newblock \bibinfo{title}{Lerch distribution based on maximum nonsymmetric
  entropy principle: Application to conway’s game of life cellular
  automaton}.
\newblock \bibinfo{journal}{Chaos, Solitons \& Fractals} \bibinfo{volume}{151},
  \bibinfo{pages}{111272}.
\newblock \URLprefix
  \url{https://www.sciencedirect.com/science/article/pii/S0960077921006263},
  \DOIprefix\doi{10.1016/j.chaos.2021.111272}.
\bibitem[{Conway(1970)}]{Conway19704}
\bibinfo{author}{Conway, J.}, \bibinfo{year}{1970}.
\newblock \bibinfo{title}{The game of life}.
\newblock \bibinfo{journal}{Scientific American} \bibinfo{volume}{223},
  \bibinfo{pages}{4}.
\newblock \URLprefix
  \url{https://www.scopus.com/inward/record.uri?eid=2-s2.0-84924933935&partnerID=40&md5=b6f44f36c7fa870548eab80106590bf9}.
  \bibinfo{note}{cited by: 27}.
\bibitem[{D'Alotto(2015)}]{d2015classification}
\bibinfo{author}{D'Alotto, L.}, \bibinfo{year}{2015}.
\newblock \bibinfo{title}{A classification of one-dimensional cellular automata
  using infinite computations}.
\newblock \bibinfo{journal}{Applied Mathematics and Computation}
  \bibinfo{volume}{255}, \bibinfo{pages}{15--24}.
\newblock \DOIprefix\doi{10.1016/j.amc.2014.10.015}.
\bibitem[{Gardner(1970)}]{gameoflife1}
\bibinfo{author}{Gardner, M.}, \bibinfo{year}{1970}.
\newblock \bibinfo{title}{Mathematical games}.
\newblock \bibinfo{journal}{Scientific American} \bibinfo{volume}{223},
  \bibinfo{pages}{120--123}.
\newblock \URLprefix \url{http://www.jstor.org/stable/24927642}.
\bibitem[{Giri et~al.(2025)Giri, Kayal and
  Contreras-Reyes}]{Giri2025Permutation}
\bibinfo{author}{Giri, R.R.}, \bibinfo{author}{Kayal, S.},
  \bibinfo{author}{Contreras-Reyes, J.E.}, \bibinfo{year}{2025}.
\newblock \bibinfo{title}{Permutation extropy: A new time series complexity
  measure}.
\newblock \bibinfo{journal}{Physica A: Statistical Mechanics and its
  Applications} \bibinfo{volume}{678}, \bibinfo{pages}{130951}.
\newblock \URLprefix
  \url{https://www.sciencedirect.com/science/article/abs/pii/S037843712500603X},
  \DOIprefix\doi{10.1016/j.physa.2025.130951}.
\bibitem[{Johnston and Greene(2021)}]{johnston2021conway}
\bibinfo{author}{Johnston, N.}, \bibinfo{author}{Greene, D.},
  \bibinfo{year}{2021}.
\newblock \bibinfo{title}{Conway's Game of Life: Mathematics and Construction}.
\newblock \bibinfo{publisher}{Self-published}.
\newblock \URLprefix \url{https://conwaylife.com/book/}.
\bibitem[{Kanungo(2024)}]{kanungo2024spacefillingcurves}
\bibinfo{author}{Kanungo, S.}, \bibinfo{year}{2024}.
\newblock \bibinfo{title}{Space-filling curves}.
\newblock \URLprefix \url{https://arxiv.org/abs/2501.04705},
  \href{http://arxiv.org/abs/2501.04705}{\tt arXiv:2501.04705}.
\bibitem[{Kari(2005)}]{KARI20053}
\bibinfo{author}{Kari, J.}, \bibinfo{year}{2005}.
\newblock \bibinfo{title}{Theory of cellular automata: A survey}.
\newblock \bibinfo{journal}{Theoretical Computer Science}
  \bibinfo{volume}{334}, \bibinfo{pages}{3--33}.
\newblock \URLprefix
  \url{https://www.sciencedirect.com/science/article/pii/S030439750500054X},
  \DOIprefix\doi{https://doi.org/10.1016/j.tcs.2004.11.021}.
\bibitem[{Kharazmi and Contreras-Reyes(2023)}]{Kharazmi2023deng}
\bibinfo{author}{Kharazmi, O.}, \bibinfo{author}{Contreras-Reyes, J.E.},
  \bibinfo{year}{2023}.
\newblock \bibinfo{title}{Deng-fisher information measure and its extensions:
  Application to conway’s game of life}.
\newblock \bibinfo{journal}{Chaos, Solitons \& Fractals} \bibinfo{volume}{174},
  \bibinfo{pages}{113871}.
\newblock \URLprefix
  \url{https://www.sciencedirect.com/science/article/abs/pii/S0960077923007725},
  \DOIprefix\doi{10.1016/j.chaos.2023.113871}.
\bibitem[{Langton(1990)}]{Langton1990Computation}
\bibinfo{author}{Langton, C.G.}, \bibinfo{year}{1990}.
\newblock \bibinfo{title}{Computation at the edge of chaos: Phase transitions
  and emergent computation}.
\newblock \bibinfo{journal}{Physica D: Nonlinear Phenomena}
  \bibinfo{volume}{42}, \bibinfo{pages}{12--37}.
\newblock \DOIprefix\doi{10.1016/0167-2789(90)90064-V}.
\bibitem[{Leyva et~al.(2022)Leyva, Martinez, Masoller, Rosso and
  Zanin}]{Leyva_2022}
\bibinfo{author}{Leyva, I.}, \bibinfo{author}{Martinez, J.},
  \bibinfo{author}{Masoller, C.}, \bibinfo{author}{Rosso, O.A.},
  \bibinfo{author}{Zanin, M.}, \bibinfo{year}{2022}.
\newblock \bibinfo{title}{20 years of ordinal patterns: Perspectives and
  challenges}.
\newblock \bibinfo{journal}{{EPL}} \bibinfo{volume}{138},
  \bibinfo{pages}{31001}.
\newblock \URLprefix \url{https://doi.org},
  \DOIprefix\doi{10.1209/0295-5075/ac6a72}.
\bibitem[{Liu(2006)}]{Liu2008}
\bibinfo{author}{Liu, C.}, \bibinfo{year}{2006}.
\newblock \bibinfo{title}{Maximal nonsymmetric entropy leads naturally to
  zipf's law}.
\newblock \URLprefix \url{https://arxiv.org/abs/cond-mat/0609344},
  \href{http://arxiv.org/abs/cond-mat/0609344}{\tt arXiv:cond-mat/0609344}.
\bibitem[{Liu(2009)}]{Liu2009}
\bibinfo{author}{Liu, C.}, \bibinfo{year}{2009}.
\newblock \bibinfo{title}{Nonsymmetric entropy and maximum nonsymmetric entropy
  principle}.
\newblock \bibinfo{journal}{Chaos, Solitons \& Fractals} \bibinfo{volume}{40},
  \bibinfo{pages}{2469--2474}.
\newblock \URLprefix
  \url{https://www.sciencedirect.com/science/article/abs/pii/S0960077907009216},
  \DOIprefix\doi{10.1016/j.chaos.2007.10.050}.
\bibitem[{Macedo and Chaimowicz(2017)}]{Macedo2017Improving}
\bibinfo{author}{Macedo, Y.P.A.}, \bibinfo{author}{Chaimowicz, L.},
  \bibinfo{year}{2017}.
\newblock \bibinfo{title}{Improving procedural 2d map generation based on
  multi-layered cellular automata and hilbert curves}, in:
  \bibinfo{booktitle}{16th Brazilian Symposium on Computer Games and Digital
  Entertainment (SBGames)}, \bibinfo{publisher}{IEEE},
  \bibinfo{address}{Curitiba, Brazil}. pp. \bibinfo{pages}{21--28}.
\newblock \bibinfo{note}{Available at
  \url{https://sbgames.org/sbgames2017/papers/ComputacaoFull/175350.pdf}}.
\bibitem[{McCrum and Kee(2024)}]{ConwaysGameo2024l}
\bibinfo{author}{McCrum, J.}, \bibinfo{author}{Kee, T.P.},
  \bibinfo{year}{2024}.
\newblock \bibinfo{title}{{C}onway’s {G}ame of {L}ife as an analogue to a
  habitable world: Livingness beyond the biological}.
\newblock \bibinfo{journal}{Proceedings of the International Astronomical
  Union} \bibinfo{volume}{20}, \bibinfo{pages}{76--86}.
\newblock \DOIprefix\doi{10.1017/s1743921324001662}.
\bibitem[{Mitchell et~al.(1993)Mitchell, Hraber and Crutchfield}]{Mitchell1993}
\bibinfo{author}{Mitchell, M.}, \bibinfo{author}{Hraber, P.T.},
  \bibinfo{author}{Crutchfield, J.P.}, \bibinfo{year}{1993}.
\newblock \bibinfo{title}{Dynamics, computation, and the ``edge of chaos'': A
  re-examination}.
\newblock \bibinfo{journal}{Complex Systems} \bibinfo{volume}{7},
  \bibinfo{pages}{89--130}.
\bibitem[{Moon et~al.(2001)Moon, Jagadish, Faloutsos and Saltz}]{Moon2001}
\bibinfo{author}{Moon, B.}, \bibinfo{author}{Jagadish, H.V.},
  \bibinfo{author}{Faloutsos, C.}, \bibinfo{author}{Saltz, J.H.},
  \bibinfo{year}{2001}.
\newblock \bibinfo{title}{Analysis of the clustering properties of the hilbert
  space-filling curve}.
\newblock \bibinfo{journal}{IEEE Transactions on Knowledge and Data
  Engineering} \bibinfo{volume}{13}, \bibinfo{pages}{124--141}.
\newblock \DOIprefix\doi{10.1109/69.908985}.
\bibitem[{Neumann and Burks(1966)}]{neumann}
\bibinfo{author}{Neumann, J.}, \bibinfo{author}{Burks, A.},
  \bibinfo{year}{1966}.
\newblock \bibinfo{title}{Theory of self-reproducing automata}.
\newblock \bibinfo{journal}{Champaign, USA: University of Illinois Press} .
\bibitem[{Pallares Di~Nunzio et~al.(2025)Pallares Di~Nunzio, Martín~Tenti,
  Arlego, Rosso and Montani}]{PallaresDiNunzio2025}
\bibinfo{author}{Pallares Di~Nunzio, M.}, \bibinfo{author}{Martín~Tenti, J.},
  \bibinfo{author}{Arlego, M.}, \bibinfo{author}{Rosso, O.A.},
  \bibinfo{author}{Montani, F.}, \bibinfo{year}{2025}.
\newblock \bibinfo{title}{Exploring the role of synaptic plasticity in the
  frequency-dependent complexity domain}.
\newblock \bibinfo{journal}{Chaos: An Interdisciplinary Journal of Nonlinear
  Science} \bibinfo{volume}{35}, \bibinfo{pages}{023138}.
\newblock \URLprefix \url{https://doi.org/10.1063/5.0239820},
  \DOIprefix\doi{10.1063/5.0239820}.
\bibitem[{Pan(2010)}]{pan2010emergencesymmetrynewtype}
\bibinfo{author}{Pan, Z.}, \bibinfo{year}{2010}.
\newblock \bibinfo{title}{Emergence from symmetry: A new type of cellular
  automata}.
\newblock \URLprefix \url{https://arxiv.org/abs/1003.3394},
  \href{http://arxiv.org/abs/1003.3394}{\tt arXiv:1003.3394}.
\bibitem[{Quintero-Quiroz et~al.(2015)Quintero-Quiroz, Pigolotti, Torrent and
  Masoller}]{QuinteroQuiroz2015}
\bibinfo{author}{Quintero-Quiroz, C.}, \bibinfo{author}{Pigolotti, S.},
  \bibinfo{author}{Torrent, M.C.}, \bibinfo{author}{Masoller, C.},
  \bibinfo{year}{2015}.
\newblock \bibinfo{title}{Numerical and experimental study of the effects of
  noise on the permutation entropy}.
\newblock \bibinfo{journal}{New Journal of Physics} \bibinfo{volume}{17},
  \bibinfo{pages}{093002}.
\newblock \URLprefix
  \url{https://iopscience.iop.org/article/10.1088/1367-2630/17/9/093002},
  \DOIprefix\doi{10.1088/1367-2630/17/9/093002}.
\bibitem[{Rendell(2011)}]{rendell2011turing}
\bibinfo{author}{Rendell, P.}, \bibinfo{year}{2011}.
\newblock \bibinfo{title}{A universal turing machine in conway's game of life},
  in: \bibinfo{booktitle}{2011 International Conference on High Performance
  Computing \& Simulation}, pp. \bibinfo{pages}{764--772}.
\newblock \DOIprefix\doi{10.1109/HPCSim.2011.5999906}.
\bibitem[{Rey et~al.(2025)Rey, Frery and Gambini}]{packageROrd}
\bibinfo{author}{Rey, A.A.}, \bibinfo{author}{Frery, A.C.},
  \bibinfo{author}{Gambini, J.}, \bibinfo{year}{2025}.
\newblock \bibinfo{title}{Asymptotic distribution of the statistical complexity
  under the multinomial law}.
\newblock \bibinfo{journal}{Chaos, Solitons \& Fractals} \bibinfo{volume}{193},
  \bibinfo{pages}{116085}.
\newblock \DOIprefix\doi{10.1016/j.chaos.2025.116085}.
\bibitem[{Rollier et~al.(2025)Rollier, Zielinski, Daly, Bruno and
  Baetens}]{Rollier_2025}
\bibinfo{author}{Rollier, M.}, \bibinfo{author}{Zielinski, K.M.},
  \bibinfo{author}{Daly, A.J.}, \bibinfo{author}{Bruno, O.M.},
  \bibinfo{author}{Baetens, J.M.}, \bibinfo{year}{2025}.
\newblock \bibinfo{title}{A comprehensive taxonomy of cellular automata}.
\newblock \bibinfo{journal}{Communications in Nonlinear Science and Numerical
  Simulation} \bibinfo{volume}{140}, \bibinfo{pages}{108362}.
\newblock \URLprefix \url{http://dx.doi.org/10.1016/j.cnsns.2024.108362},
  \DOIprefix\doi{10.1016/j.cnsns.2024.108362}.
\bibitem[{Rosso et~al.(2007)Rosso, Larrondo, Martin, Plastino and
  Fuentes}]{Rosso2007}
\bibinfo{author}{Rosso, O.A.}, \bibinfo{author}{Larrondo, H.A.},
  \bibinfo{author}{Martin, M.T.}, \bibinfo{author}{Plastino, A.},
  \bibinfo{author}{Fuentes, M.A.}, \bibinfo{year}{2007}.
\newblock \bibinfo{title}{Distinguishing noise from chaos}.
\newblock \bibinfo{journal}{Phys. Rev. Lett.} \bibinfo{volume}{99},
  \bibinfo{pages}{154102}.
\newblock \URLprefix
  \url{https://link.aps.org/doi/10.1103/PhysRevLett.99.154102},
  \DOIprefix\doi{10.1103/PhysRevLett.99.154102}.
\bibitem[{Speck and Glauber(1978)}]{Speck1978}
\bibinfo{author}{Speck, D.R.}, \bibinfo{author}{Glauber, R.J.},
  \bibinfo{year}{1978}.
\newblock \bibinfo{title}{Statistical mechanics of a dynamical system based on
  conway's game of life}.
\newblock \bibinfo{journal}{Journal of Statistical Physics}
  \bibinfo{volume}{19}, \bibinfo{pages}{215--249}.
\newblock \DOIprefix\doi{10.1007/BF01011727}.
\bibitem[{Staniek and Lehnertz(2007)}]{staniek2007parameter}
\bibinfo{author}{Staniek, M.}, \bibinfo{author}{Lehnertz, K.},
  \bibinfo{year}{2007}.
\newblock \bibinfo{title}{Parameter selection for permutation entropy
  measurements}.
\newblock \bibinfo{journal}{International Journal of Bifurcation and Chaos}
  \bibinfo{volume}{17}, \bibinfo{pages}{2273--2282}.
\bibitem[{Wolfram(1983)}]{Wolfram1983Statistical}
\bibinfo{author}{Wolfram, S.}, \bibinfo{year}{1983}.
\newblock \bibinfo{title}{Statistical mechanics of cellular automata}.
\newblock \bibinfo{journal}{Rev. Mod. Phys.} \bibinfo{volume}{55},
  \bibinfo{pages}{601--644}.
\newblock \URLprefix \url{https://link.aps.org/doi/10.1103/RevModPhys.55.601},
  \DOIprefix\doi{10.1103/RevModPhys.55.601}.
\bibitem[{Wolfram(1984)}]{WOLFRAM19841}
\bibinfo{author}{Wolfram, S.}, \bibinfo{year}{1984}.
\newblock \bibinfo{title}{Universality and complexity in cellular automata}.
\newblock \bibinfo{journal}{Physica D: Nonlinear Phenomena}
  \bibinfo{volume}{10}, \bibinfo{pages}{1--35}.
\newblock \URLprefix
  \url{https://www.sciencedirect.com/science/article/pii/0167278984902458},
  \DOIprefix\doi{https://doi.org/10.1016/0167-2789(84)90245-8}.
\bibitem[{Wolfram(2002)}]{Wolfram2002}
\bibinfo{author}{Wolfram, S.}, \bibinfo{year}{2002}.
\newblock \bibinfo{title}{A New Kind of Science}.
\newblock \bibinfo{publisher}{Wolfram Media}, \bibinfo{address}{Champaign, IL}.
\newblock \URLprefix \url{https://www.wolframscience.com}.

\end{thebibliography}

\end{document}